\documentclass[structabstract]{aa}  
\usepackage{subfig}
\usepackage{float}
 
\usepackage{graphicx}
\usepackage{lscape}
\usepackage{epsfig}
\usepackage{txfonts}
\usepackage{natbib}
\usepackage{times}
\begin{document}
   \title{Fossil groups origins II.\\
 Unveiling the formation of the brightest group galaxies through their scaling relations}
   \subtitle{}

   \author{J. M\'endez-Abreu
          \inst{1,2}
        \and
        J. A. L. Aguerri\inst{1,2}
        \and
        R. Barrena\inst{1,2}
          \and
        R. S\'anchez-Janssen\inst{3}
          \and
W. Boschin\inst{4}
\and\\
N. Castro-Rodriguez\inst{1,2}
\and
E. M. Corsini\inst{5}
\and
C. del Burgo\inst{6}
\and
E. D'Onghia\inst{7}
\and
M. Girardi\inst{8,9}
\and
J. Iglesias-P\'aramo\inst{10,11}
\and\\
N. Napolitano\inst{12}
\and
J. M. Vilchez\inst{10}
\and
S. Zarattini\inst{1,2,5}
          }

\institute{Instituto Astrof\'\i sico de
           Canarias, C/ V\'ia L\'actea s/n, 38200 La Laguna, Spain
         \and Departamento de Astrof\'\i sica, Universidad de La Laguna, 
              C/ Astrof\'isico Francisco S\'anchez, 38205 La Laguna, Spain \\
         \email{jairo@iac.es} 
         \and European Southern Observatory, Alonso de C\'ordova 3107, Vitacura, Santiago, Chile
         \and Fundaci\'on Galileo Galilei-INAF, Rambla Jos\'e Ana Fern\'andez P\'erez 7, 38712 Bre\~na Baja, Spain 
         \and Dipartimento di Astronomia, Universit\`a di Padova, vicolo dell'Osservatorio 3, 35122 Padova, Italy 
         \and UNINOVA/CA3, Campus da FCT/UNL, Quinta da Torre, 2825-149 Caparica, Portugal 
         \and Harvard-Smithsonian Center for Astrophysics, 60 Garden Street, Cambridge, MA 02138, USA 
         \and Dipartimento di Fisica-Sezione Astronomia, Universit\`a degli Studi di Trieste, via Tiepolo 11, 34143 Trieste, Italy 
         \and INAF-Osservatorio Astronomico di Trieste, via Tiepolo 11, 34143 Trieste, Italy
         \and Instituto de Astrof\'{\i}sica de Andalucia-C.S.I.C., 18008 Granada, Spain  
         \and Centro Astron\'omico Hispano Alem\'an, C/ Jes\'us Durb\'an Rem\'on 2-2, 04004 Almer\'{\i}a, Spain 
         \and INAF-Osservatorio Astronomico di Capodimonte, Salita Moiariello 16, 80131 Napoli, Italy 
         }

   \date{Received September 15, 1996; accepted March 16, 1997}

 
  \abstract
%
{Fossil  systems are  galaxy  associations dominated  by a  relatively
  isolated, bright elliptical galaxy, surrounded by a group of smaller
  galaxies   lacking  $L^{*}$   objects.  Because   of   this  extreme
  environment, fossil groups (FGs) are ideal laboratories to study the
  mass assembly of brightest group galaxies (BGGs).}
%
%
{We analyzed  the near-infrared photometric  and structural properties
  of a sample of 20 BGGs  present in FGs in order to better understand
  their formation  mechanisms. This represent  the largest sample
    studied to date.}
%
%
{$K_s$-band deep  images were used to study  the structural properties
  of our  sample galaxies.  Their  surface-brightness distribution was
  fitted to a S\'ersic profile using the GASP2D algorithm.  Then,
    the standard scaling relations were derived for the first time for
    these galaxies and compared  with those of normal ellipticals and
  brightest cluster galaxies in non-fossil systems.}
%
%
{The  BGGs presented  in this  study represent  a subset  of  the most
  massive galaxies  in the Universe.  We found  that their ellipticity
  profiles are continuously increasing with the galactocentric radius.
  Our fossil  BCGs follow closely  the fundamental plane  described by
  normal  ellipticals.   However,  they  depart from  both  the  $\log
  \sigma_0$ vs.  $\log L_{K_{s}}$ and $\log r_{\rm e}$ vs.  $\log L_{K_{s}}$
  relations described  by intermediate mass  ellipticals.  This occurs
  in the sense  that our BGGs have larger  effective radii and smaller
  velocity  dispersions than  those predicted  by these  relations. We
  also found that more elliptical galaxies systematically deviate from
  the previous relations while more  rounder object do not. No similar
  correlation was found with the S\'ersic index.}
%
%
   {The derived scaling  relations can be interpreted in  terms of the
     formation  scenario of  the BGGs.   Because our  BGGs  follow the
     fundamental plane tilt but  they have larger effective radii than
     expected for intermediate mass  ellipticals, we suggest that they
     only went through dissipational mergers in a early stage of their
     evolution  and then  assembled  the bulk  of  their mass  through
     subsequent dry mergers, contrary  to previous claims that BGGs in
     FGs were formed mainly by the merging of gas-rich galaxies.}


\authorrunning{J. M\'endez-Abreu et al.}
\titlerunning{Fossil groups origins. II.}

   \maketitle
%

\section{Introduction}
\label{sec:intro}


Fossil  groups  (FGs)  are  observationally defined  as  X-ray  bright
systems ($L_X  > 10^{42} h^{-2}_{50}$ erg s$^{-1}$),  dominated in the
optical  by a  giant  elliptical galaxy  at  the center,  and with  an
$R$-band difference in  the absolute magnitude of $\Delta  M_{12} > 2$
\citep{jones03} between the two brightest galaxies located within half
the virial radius  of the system.  These relatively  rare systems were
first identified  by \citet{ponman94} and  were the target  of several
studies            in             the            last            years
\citep{vikhlinin99,jones03,mendesdeoliveira06,cypriano06,khosroshahi06,mendesdeoliveira09,zibetti09,democles10,aguerri11}.

Observationally,  these systems are  massive galaxy  associations with
typical masses of rich groups  or poor clusters.  They follow also the
scaling  relations of  groups  or clusters  such  as X-ray  luminosity
($L_X$) vs.  gas temperature ($T_X$), total mass ($M$) vs.  $T_X$, gas
entropy  vs.    $T_X$  ,   $L_X$  vs.   cluster   velocity  dispersion
($\sigma_{\rm   cl}$),   optical   luminosity  ($L_{\rm   opt}$)   vs.
$\sigma_{\rm  cl}$,  $\sigma_{\rm  cl}$  vs.   $T_X$  ,  gas  fraction
($f_{\rm   gas}$)   vs.    $T_X$  \citep{khosroshahi07,sun09}.    Some
differences  have been  found  in the  optical  vs.  X-ray  luminosity
($L_{\rm opt}$-$L_X$)  relation \citep{khosroshahi07} even  if this is
still controversial  \citep{voevodkin10}.  Detailed X-ray observations
of some FGs  also indicate that these systems  were assembled at early
epochs  in high  centrally concentrated  dark matter  (DM)  halos with
large    mass-to-light    ratios    ($M/L$)    \citep[][]{democles10}.
Nevertheless,  they do  not show  cooling cores  as those  detected in
galaxy  clusters, which points  toward the  presence of  other heating
mechanisms,    like   active    galactic    nuclei   (AGN)    feedback
\citep{sun04,khosroshahi04,khosroshahi06,mendesdeoliveira09}.

From numerical simulations, and due to the interesting nature of these
systems, their formation  scenarios have been matter of  debate in the
last                                                              years
\citep{donghia05,sommerlarsen06,vonbenda-beckmann08,romeo08,diazgimenez11,cui11}.
One scenario  attributes their properties to  their dynamical history,
implying  that fossil systems  are formed  if a  group/cluster remains
undisturbed for a  long period of time.  Therefore,  if fossil systems
were assembled in an early epoch of the Universe throughout a fast and
efficient process  of merging, they  should had time enough  for their
$L^{*}$  galaxies  to  merge,  thus  producing the  observed  lack  of
intermediate-luminosity galaxies  and the large  magnitude gap between
the  brightest   and  the  second   brightest  galaxy  of   the  group
\citep{donghia05,  vonbenda-beckmann08}.   An alternative  explanation
for FG formation is that  such systems are “failed” groups that suffer
from   a  lack   of  $L^{*}$   galaxies  as   an  accident   of  birth
\citep{mulchaeyzabludoff99}. On  the other hand,  some authors suggest
that FGs might represent a  transitory phase in the cluster life, with
an absence of significant mergers  for a long time, enough for cluster
relaxation \citep{vonbenda-beckmann08,cui11}.


The merger  history underlying the  origin of the central  galaxies in
FGs continue to be discussed  in the literature. Their brightest group
galaxies  (BGGs) are  among the  most  massive galaxies  known in  the
Universe.   Originally, they were  thought to  be possible  results of
group     evolution     as     driven    by     dynamical     friction
\citep{ponman94,jones03}.    Recent   N-body  simulations   and
  semi-analytic  calculations have  suggested  that BGGs  formed in  a
  short time-scale as consequence of mergers with low angular momentum
  \citep{sommerlarsen06}, pushing dynamical  friction to a less likely
  scenario for the evolution of BGGs.

Observations  show   that  BGGs  have   also  different  observational
properties than other bright  elliptical galaxies. In particular, they
present disky isophotes in the center and their luminosities correlate
with  the  velocity  dispersions   of  the  groups.   These  different
properties   suggest  a  different   formation  scenario   for  bright
ellipticals  in fossil  and non-fossil  systems \citep{khosroshahi06}.
Whereas bright ellipticals in FGs  would grow out of gas-rich mergers,
giant ellipticals in non-fossil systems would suffer more dry mergers.
However,  neither  recent  samples  of  BGGs  \citep{labarbera09}  nor
numerical  simulations  \citep{diazgimenez08}  show these  differences
\citep{labarbera09}.  All previous results  on fossil systems have the
drawback  that they  were obtained  using small  samples of  FGs. This
could be the reason of  some contradictory findings found by different
studies.  The  lack of  a large and  homogeneous statistical  study of
this kind of systems makes the previous results not conclusive.


In  the  framework  of   the  Fossil  Groups  Origins  (FOGO)  project
\citep{aguerri11},   which   aims   to   carry   out   a   systematic,
multiwavelength study  of a sample of  34 FGs selected  from the Sloan
Digital Sky Survey  \citep[SDSS;][]{santos07}, near-infrared images in
the $K_s$-band were  taken for the central regions  of 17 groups using
the Long-slit Intermediate Resolution Infrared Spectrograph (LIRIS) at
the 4.2-m  William Herschel Telescope  (WHT). This  represent the
  largest  sample  of BGGs  studied  to  date  at these  wavelengths.
Near-infrared   observations  have  the   advantage  of   mapping  the
distribution  of  the mass-carrying  evolved  stars  and diminish  the
influence of dust. In this  work, we analyze in detail the photometric
properties of  the BGGs present on  these FGs to  perform a systematic
study  of  their  structural   properties  and  shed  light  on  their
formation/evolution scenarios.


The  remainder  of  this  paper   is  as  follows:  the  data  sample,
observations,     and    data     reduction    are     summarized    in
Sect.~\ref{sec:data}.   In  Sect.~\ref{sec:surface}  we  describe  the
surface   photometry  and   photometric  decomposition.   The  scaling
relations are presented  in Sect.~\ref{sec:scaling}. The discussion of
the     results      and     conclusions     are      presented     in
Sects.~\ref{sec:discussion}, and \ref{sec:conclusions}, respectively.

Throughout  this paper we  assume a  flat cosmology  with $\Omega_{\rm
  M}=0.3$, $\Omega_{\Lambda}=0.7$,  and a Hubble  parameter H$_{0}=70$
km s$^{-1}$ Mpc$^{-1}$.

\section{Data sample, observations, and data reduction}
\label{sec:data}

\subsection{LIRIS data}
\label{sec:liris}

Near-infrared imaging  of a sample of  17 FG systems  belonging to the
FOGO project was carried out at  the WHT in January 2010.  We observed
in  the  $K_s$-band  using  the   LIRIS  imaging  mode.   LIRIS  is  a
near-infrared  (0.9-2.4 $\mu$m)  instrument with  an optical  system
based on  a classical camera design  \citep{manchado04}.  The detector
is a  Hawaii 1024$\times$1024  HgCdTe array operating  at 70 K  with a
pixel scale of 0\farcs25 pixel$^{-1}$.

We exposed a total of 3000 s per field, developing several cycles of a
5 points dithering  for each target.  The exposure  time of individual
frames  was 10-12  s. The  dithering cycling  procedure allowed  us to
estimate and then correct the  background contributions of the sky and
its  significant oscillations.   In addition,  the dithering  and flat
frames helped  us to clean cosmetic  effects of the  CCDs. After these
corrections,  we obtained  maximum variations  of about  0.4\%  in the
background across the full image.

Data reduction was carried out under IRAF\footnote{IRAF is distributed
  by the National Optical  Astronomy Observatories, which are operated
  by the Association of  Universities for Research in Astronomy, Inc.,
  under cooperative  agreement with the  National Science Foundation.}
environment and LIRIS specific packages.  In order to keep the quality
of  the point spread  function (PSF)  across the  whole field  of view
after stacking individual frames, the geometry effect of the distorted
field was  corrected.  In  the combined images  the PSF  is consistent
with that of the single ones. This  is true even close to the edges of
the field.

Our  observations were  carried out  under photometric  conditions and
within  a seeing  range  of 0\farcs6-1\farcs1  (FWHM)  as measured  by
fitting a  Moffat PSF to field  stars in our  images.  The photometric
calibration  was  performed  by  observing  United  Kingdom  Infra-Red
Telescope       (UKIRT)       faint      near-infrared       standards
\citep{casalihawarden92}.  We have checked our photometric calibration
taking the Two Micron All  Sky Survey (2MASS) photometry as reference.
Our  zero-point estimations  were  in agreement  with  those of  2MASS
within $\pm 0.02$ mag.

Finally we  identified galaxies in our $K_s$-band  images and measured
their magnitudes with  the SExtractor package \citep{bertinarnouts96}.
Objects  were identified imposing  that they  cover a  certain minimum
area and  have a number counts  above a limiting  threshold taking the
local sky background as a reference. The limiting size and flux were 9
pixels and 1.5 standard deviation of the sky counts, respectively. The
selected limiting  size corresponds to an apparent  size of 0\farcs75,
which  is about  the typical  seeing FWHM  obtained from  our combined
images. We performed careful visual inspections of the frames in order
to deal with the best combination of the above parameters that removes
spurious objects from the catalogues.

   \begin{figure}[!t]
   \centering
   \includegraphics[width=0.49\textwidth]{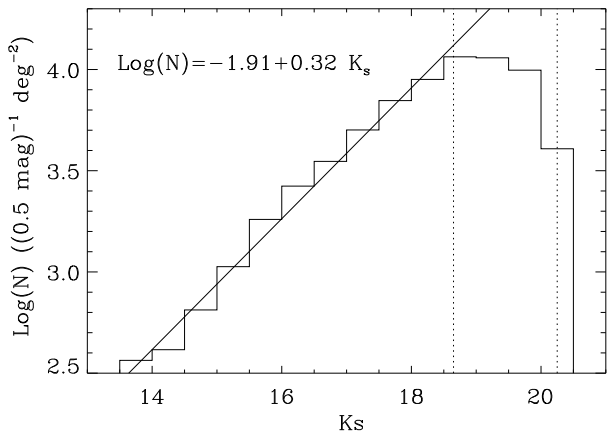}
   \caption{$K_s$-band counts per square  degree per 0.5 magnitude bin
     for  the fields observed  with LIRIS.   The continuous  curve and
     vertical dotted  lines represent the logarithmic  function fit to
     the histogram, completeness, and limit magnitudes, respectively.}
              \label{fig:counts}%
    \end{figure}

Figure  1 shows  the $K_s$-band  counts for  the fields  observed with
LIRIS.  The  counts follow the  count-magnitude relation expected  for a
homogeneous   galaxy  distribution  in   a  universe   with  Euclidean
geometry.  The completeness was  computed as  the magnitude  where the
measured counts  are less than 10\%  with respect to  the fit, whereas
the limit  magnitude was assumed to  be that of  the faintest detected
objects.   For  our  sample,  the completeness  and  limit  magnitudes
correspond to  targets with signal-to-noise  ratio $S/N$=5 and  3, and
are  $K_s=18.7$ and  20.3, respectively.   The characteristics  of the
observed sample are listed in Table \ref{tab:observations}.

\subsection{Additional data}
\subsubsection{UKIDSS data}
\label{sec:UKIDSS}

Since we were  not able to observe  the whole sample of 34  FGs of the
FOGO sample,  we carried out a  search in the UKIRT  Infrared Deep Sky
Survey  (UKIDSS)   eighth  data  release  database   to  complete  our
sample. We found 7 galaxies out of 17 already observed with LIRIS.  In
addition, 5 new groups not previously observed with LIRIS were found.

   \begin{figure*}
   \centering
   \includegraphics[width=0.8\textwidth]{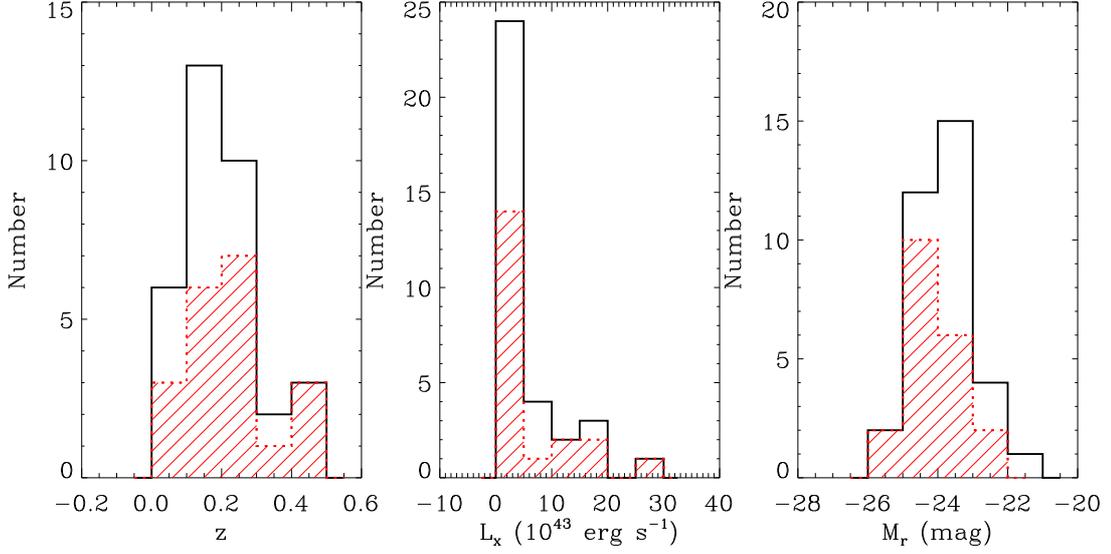}
   \caption{Distribution of  redshift (left panel),  X-ray luminosity
     (middle panel), and $r$-band absolute magnitude (right panel) for
     the FOGO FGs  (solid black line) and of FGs  studied in this work
     (red dotted line).}
              \label{fig:comp_samples}%
    \end{figure*}

Since the quality and depth of  the UKIDSS data is clearly poorer than
those of images obtained using LIRIS,  we analyzed the images of the 7
common BGGs (see Table.~\ref{tab:observations})
%
%
to  understand  whether  the  UKIDSS  data  could  be  useful  to  our
scientific  purpose.  We  perform a  photometric decomposition  of the
surface-brightness  distribution of  the BGGs  measured on  the UKIDSS
images as we  did for the LIRIS data  (see Sect.~\ref{sec:photdec} for
details). By  comparing the total magnitudes obtained  from both fits,
we found that only for 4 BGGs (see Table.~\ref{tab:observations})
%
%
the results are comparable within 0.2 mag. After analyzing the results
of the  fits we concluded  that in order  to obtain a reliable  fit we
need a suitable combination  of surface-brightness dynamical range and
S\'ersic index, i.e., we found that galaxies with S\'ersic index $n>3$
need at least a range of  6 mag arcsec$^{-2}$ to be well fitted within
the observational errors;  instead, for $n\leq3$ we need  only a range
of 4  mag arcsec$^{-2}$.  Applying these  criteria to the  5 images of
the  BGGs without  counterpart in  LIRIS  we found  that three  groups
(SDSSJ154855.85+085044.3,                      SDSSJ161431.10+264350.3,
SDSSJ225630.04-003210.8)  satisfy the conditions  and only  these were
included in the analysis.

It is  worth notice that we are not  covering the complete sample
  of  34 FGs  comprised by  the FOGO  project.  This  subsample  of 20
  groups is missing mainly the fainter  end of the FOGO sample in both
  optical  magnitude   of  the  BGG   and  X-ray  luminosity   of  the
  groups. However, this  bias does not affect the  results obtained in
  this   article.   In  Fig.    \ref{fig:comp_samples}  we   show  the
  distribution of  the complete FOGO  sample and that studied  in this
  paper as  function of the  redshift, X-ray luminosity,  and $r-$band
  absolute magnitude.

\subsubsection{SDSS data}
\label{sec:spectro}

Since  all  our BGGs  were  observed  spectroscopically  by the  Sloan
Digital Sky  Survey (SDSS), we  searched for the presence  of velocity
dispersion  measurements in the SDSS database.  We found the
stellar velocity  dispersions for the whole sample  using the SDSS-III
eighth  data  release  \citep{sdssiii11}.   However,  since  the  SDSS
spectroscopy was  performed using fibers  with $3''$ of  diameter, the
measurement of  the velocity dispersion  could be contaminated  by the
presence of another  objects inside this radius.  For  this reason, we
checked visually the  galaxies in order to take  into account for this
effect and we did not use  the velocity dispersions of the BGGs in the
SDSSJ084257.55+362159.2,     and    SDSSJ133559.98-033129.1    groups.
Therefore,  our  final  sample  of  velocity  dispersion  includes  18
galaxies.  The  velocity dispersions  were corrected for  the aperture
size  effect  following the  prescription  by \citet{jorgensen95}  and
adopting a fixed physical aperture of $r_{\rm e}$/8.  The measurements
after aperture corrections are given in Table \ref{tab:observations}.

\begin{table*}[!ht]
\caption{\label{tab:observations} Characteristics of our sample of FGs
  and BGGs.}  
\centering
\begin{tabular}{l c c c c c c}
\hline\hline
\multicolumn{1}{c}{Group} &
\multicolumn{1}{c}{$L_x$}&
\multicolumn{1}{c}{$z$}&
\multicolumn{1}{c}{m$_{r}$}&
\multicolumn{1}{c}{$\sigma_0$}&
\multicolumn{1}{c}{Source}&
\multicolumn{1}{c}{FWHM}\\
\multicolumn{1}{c}{}&
\multicolumn{1}{c}{($\times$10$^{43}$ erg/s)}&
\multicolumn{1}{c}{}&
\multicolumn{1}{c}{(mag)}&
\multicolumn{1}{c}{(km/s)}&
\multicolumn{1}{c}{}&
\multicolumn{1}{c}{(arcsec)}\\
\multicolumn{1}{c}{(1)}&
\multicolumn{1}{c}{(2)}&
\multicolumn{1}{c}{(3)}&
\multicolumn{1}{c}{(4)}&
\multicolumn{1}{c}{(5)}&
\multicolumn{1}{c}{(6)}&
\multicolumn{1}{c}{(7)}\\
\hline
SDSSJ015021.27-100530.5  &  14.40  & 0.365  & 17.26 & 341$\pm$44  & L & 0.7 \\
SDSSJ015241.95+010025.5  &  15.10  & 0.230  & 15.72 & 317$\pm$16  & L,U& 1.1 \\
SDSSJ075244.19+455657.3  &  0.562  & 0.052  & 14.46 & 221$\pm$5   & L & 0.8 \\
SDSSJ080730.75+340041.6  &  4.210  & 0.208  & 16.38 & 263$\pm$21  & L & 0.7 \\
SDSSJ084257.55+362159.2  &  29.50  & 0.282  & 16.79 &    ...      & L & 0.8 \\
SDSSJ084449.07+425642.1  &  0.211  & 0.054  & 14.08 & 157$\pm$4   & L & 0.7 \\
SDSSJ090303.18+273929.3  &  17.40  & 0.489  & 18.06 & 273$\pm$54  & L,U$^*$& 0.7 \\
SDSSJ094829.04+495506.7  &  6.260  & 0.409  & 18.21 & 270$\pm$30  & L & 0.6 \\
SDSSJ104302.57+005418.2  &  4.990  & 0.125  & 15.98 & 235$\pm$8   & L,U$^*$& 0.9 \\
SDSSJ105452.03+552112.5  &  11.70  & 0.468  & 17.69 & 365$\pm$61  & L & 0.7 \\
SDSSJ111439.76+403735.1  &  4.190  & 0.202  & 17.14 & 218$\pm$11  & L & 0.6 \\
SDSSJ112155.27+104923.2  &  3.850  & 0.240  & 16.97 & 251$\pm$21  & L,U$^*$& 0.7 \\
SDSSJ114128.29+055829.5  &  2.190  & 0.188  & 16.03 & 309$\pm$28  & L,U& 0.8 \\
SDSSJ114647.57+095228.1  &  4.930  & 0.221  & 16.36 & 276$\pm$18  & L,U$^*$& 0.6 \\
SDSSJ124742.07+413137.6  &  0.625  & 0.155  & 15.88 & 240$\pm$13  & L & 0.6 \\
SDSSJ133559.98-033129.1  &  3.680  & 0.177  & 15.84 &   ...       & L & 0.9 \\
SDSSJ154855.85+085044.3  &  0.509  & 0.072  & 13.50 & 326$\pm$9   & U & 0.7 \\
SDSSJ161431.10+264350.3  &  2.370  & 0.184  & 15.76 & 276$\pm$15  & U & 0.7 \\
SDSSJ225630.04-003210.8  &  2.180  & 0.224  & 16.81 & 274$\pm$20  & U & 0.7 \\
SDSSJ235815.10+150543.5  &  0.926  & 0.178  & 16.08 & 297$\pm$14  & L,U& 0.8 \\
\hline
\end{tabular}
\tablefoot{Col.   (1): Group  name; Col.   (2): X-ray  luminosity from
  \citet{santos07};  Cols.   (3,   4,  5):  redshift,  $r$-band  model
  magnitude, and  aperture corrected  velocity dispersion of  the BGGs
  derived   from  SDSS;   Col.    (6):  source   of  the   photometric
  observations: L=LIRIS,  U=UKIDSS, common galaxies  with successfully
  fitted UKIDSS images are marked  with an asterisk; Col. (7): FWHM of
  the Moffat PSF fitted to the stars in the image field of view.}
\end{table*}

\section{Surface photometry}
\label{sec:surface}
\subsection{Photometric decomposition}
\label{sec:photdec}

The two-dimensional  photometric decomposition of  the sample galaxies
was    performed   using   the    GASP2D   algorithm    described   in
\citet{mendezabreu08}.  This  algorithm has been  successfully applied
to                several                galaxy                samples
\citep{morelli08,pizzella08,beifiori11,dallabonta11}.   In particular,
it has  been applied to a  large sample of  brightest cluster galaxies
(BCGs) in non-fossil systems \citep{ascaso11}

The GASP2D  algorithm is based on a  $\chi^2$ minimization. Therefore,
the choice  of the  initial trials for  free parameters is  crucial to
obtain a good  fit. To this aim, the  ellipse-averaged radial profiles
of surface  brightness, ellipticity, and position  angle were analyzed
by  following the  prescriptions  by \citet{mendezabreu08}.   Starting
from these  initial trial parameters the  different photometric models
of the  surface brightness  were fitted iteratively  by GASP2D  to the
pixels  of  the galaxy  image  to  derive  its photometric  structural
parameters.  Each  image pixel was weighted according  to the variance
of its  total observed photon counts  due to the  contribution of both
galaxy and  sky, and determined  assuming photon noise  limitation and
taking the detector read-out  noise into account.  Seeing effects were
also taken into account by  convolving the model image with a circular
Moffat  PSF  with  a  FWHM   matching  the  observed  one  (see  Table
\ref{tab:observations}).  The  convolution was performed  as a product
in Fourier domain before the least-squares minimization.

We  tested  three  different  parametric  models  to  fit  the  galaxy
surface-brightness   distribution.    The   de   Vaucouleurs   profile
\citep{devaucouleurs48}  was extensively  used to  describe  the light
distribution of  elliptical galaxies and to look  for central dominant
galaxies (cDs) in clusters by searching light excess over the model de
Vaucouleurs   profile   in   the    outer   parts   of   the   profile
\citep{matthews64,schombert87}.        The       S\'ersic      profile
\citep{sersic68} is  nowadays believed to be the  most suitable model
to describe the surface-brightness distribution of elliptical galaxies
in     a     wide     range     of     luminosities     and     masses
\citep[e.g.,][]{caon93,grahamguzman03,aguerri04,kormendy09}.    It  is
parametrized  by   $r_{\rm  e}$,  I$_{\rm   e}$,  and  $n$   which  are
respectively the  effective radius, surface brightness  at r$_{\rm e}$
and a shape parameter describing  the curvature of the radial profile.
When  $n=4$ we recover  the de  Vaucouleurs profile  and for  $n=1$ we
obtain  the exponential  one.  Figure~\ref{fig:SB_profiles}  shows the
best  fit  of the  observed  surface-brightness  radial  profile to  a
S\'ersic model  obtained for our  sample galaxies. In addition  to the
previous models, we also fitted a two-component (S\'ersic+exponential)
model to the data.  This  allowed us to fit the low surface-brightness
and     extended     component     that     BCGs     could     present
\citep{nelson02,gonzalez03,gonzalez05,seigar07,liu08,ascaso11}.      We
found  that 7  galaxies  are described  by  this two-component  model,
however, since all galaxies can be successfully modelled by a S\'ersic
profile and in order to work with a self-comparable set of structural
parameters, we decided to  use only the S\'ersic structural parameters
in this paper.  In addition, we  found that a high fraction (14 out of
20) of our sample galaxies  are compatible within the errors bars with
following a  de Vaucouleurs profile.  Following the  criteria given by
\citet{schombert87}  to identify cDs  based on  looking for  an upward
break in the surface-brightness profile with respect to the typical de
Vaucouleurs profile, we identified SDSSJ124742.07+413137.6 as the only
one candidate to  be a cD galaxy. However,  this classification should
be  interpreted  as  tentative,  since the  surface-brightness  limits
reached in  our photometry are  not directly comparable with  those of
\citet{schombert87}.

%
\begin{figure*}[!ht]
  \centering 
  \subfloat[ ][]{\includegraphics[width=\textwidth]{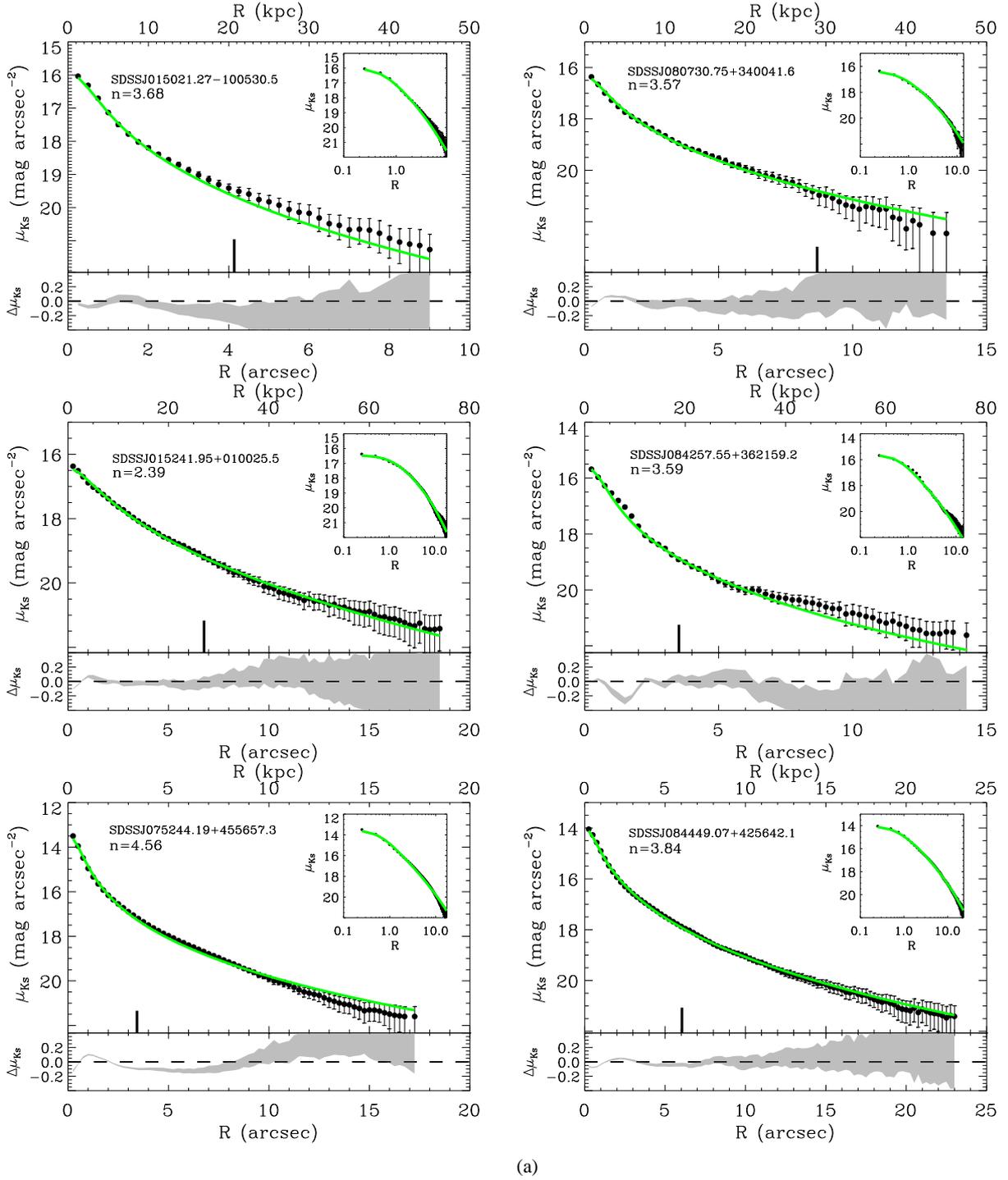}}%
   \caption{Observed  (black  points)  and  model (green  solid  line)
     surface-brightness  radial profiles of  the sample  galaxies. The
     observed radial profiles were obtained by fitting ellipses to the
     isophotes of the $K_s$-band  images.  The model profiles follow a
     S\'ersic law. These profiles  were not corrected for cosmological
     dimming, K- or evolutionary corrections. The short vertical lines
     mark the effective radius. Semi-major axes of fitted ellipses are
     given in  arcsec (bottom) and kpc  (top). The inset  in the upper
     panels  shows  the surface  brightness  in  logarithmic scale  to
     enhance  the  quality of  the  fit  in  the inner  regions.   The
     residuals of  the fit are represented  with a grey  region in the
     lower panels.}
\end{figure*}
\begin{figure*}[!ht]
  \ContinuedFloat 
  \centering 
  \subfloat[ ][]{\includegraphics[width=\textwidth]{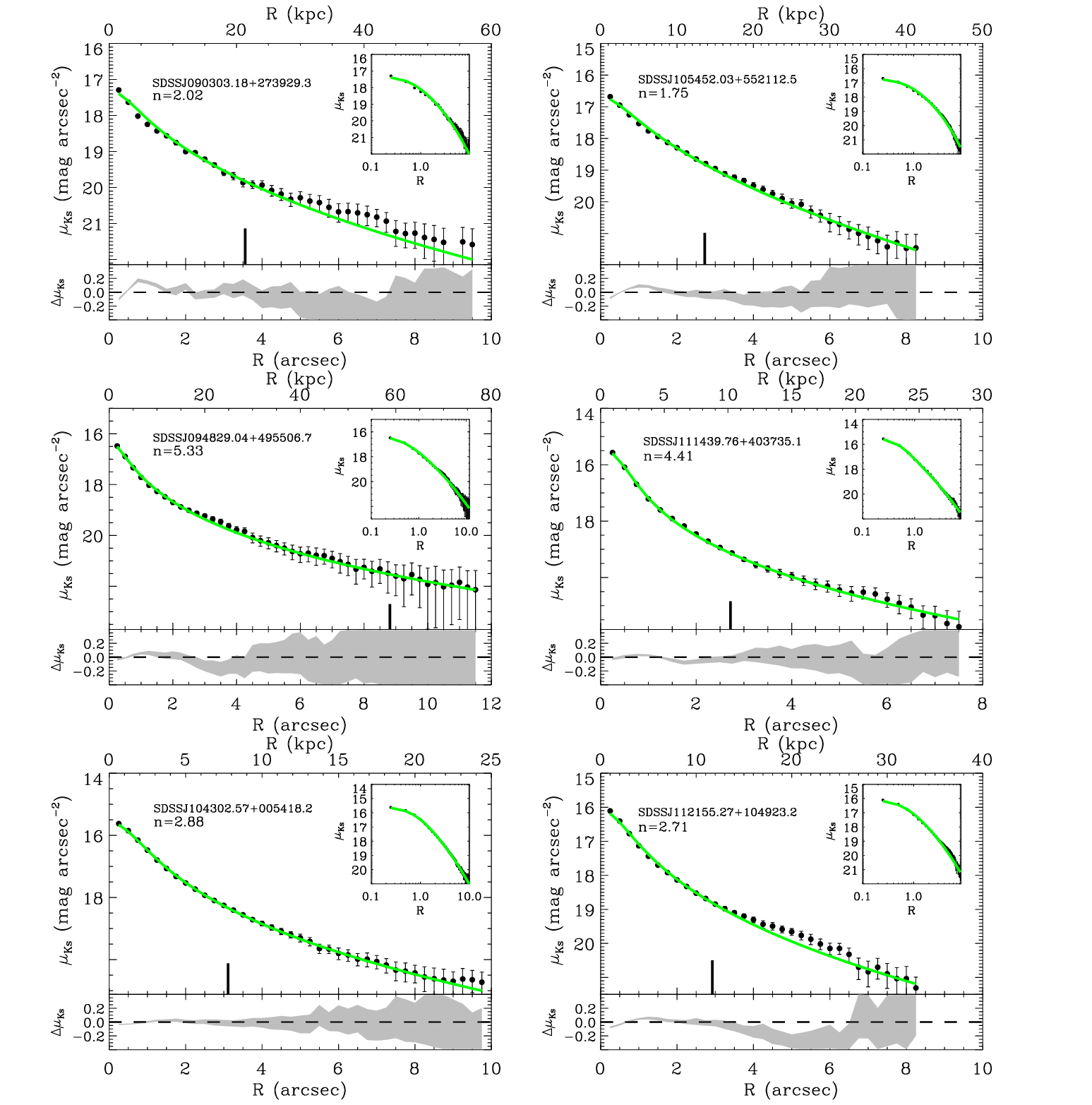}}%
  \caption[]{Continued.}
  \label{fig:cont}
\end{figure*} 
\begin{figure*}[!ht]
  \ContinuedFloat 
  \centering 
  \subfloat[ ][]{\includegraphics[width=\textwidth]{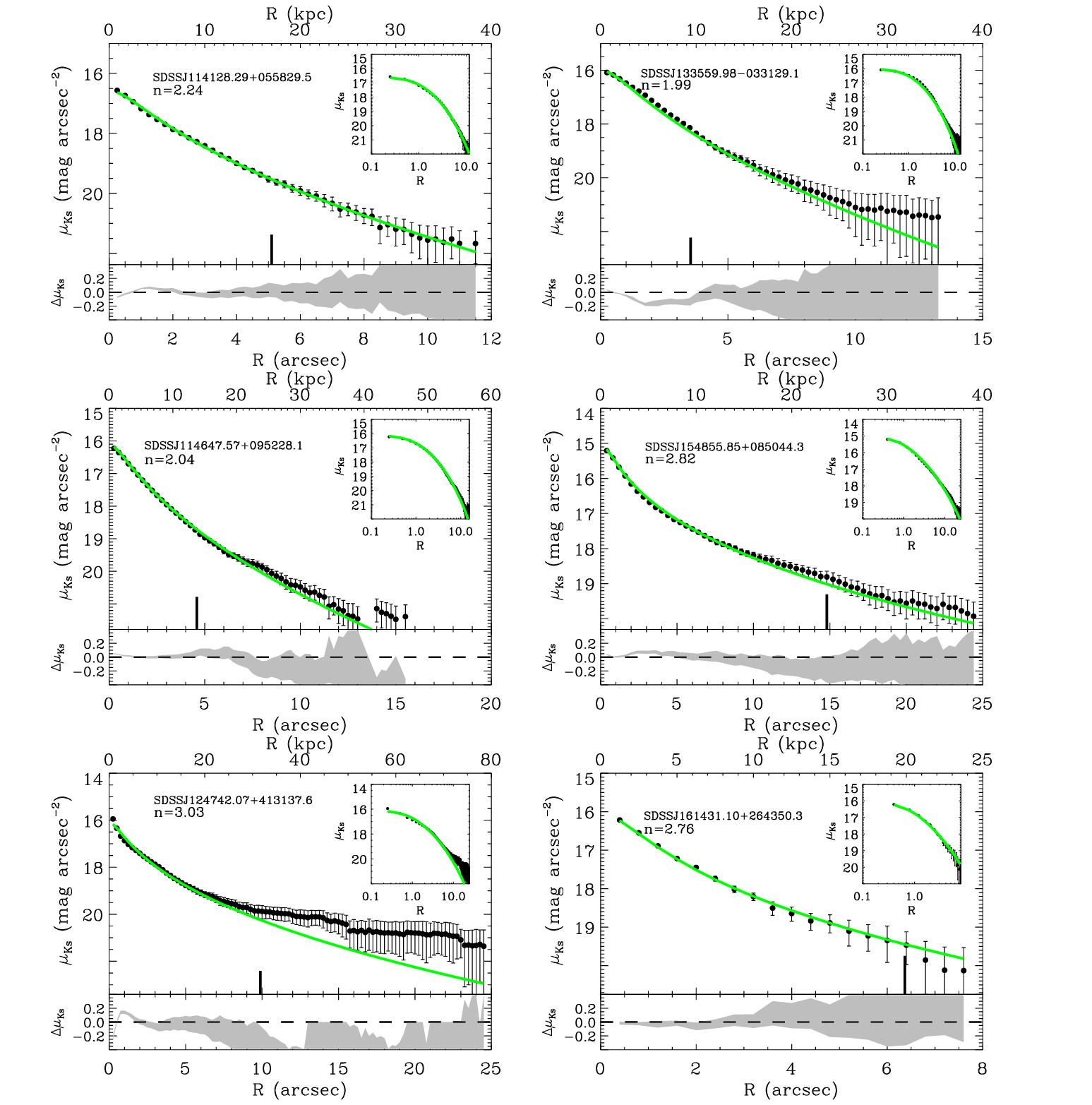}}%
  \caption[]{Continued.}
  \label{fig:cont}
\end{figure*} 
\begin{figure*}[!ht]
  \ContinuedFloat 
  \centering 
  \subfloat[ ][]{\includegraphics[width=\textwidth,bb=54 680 558 870]{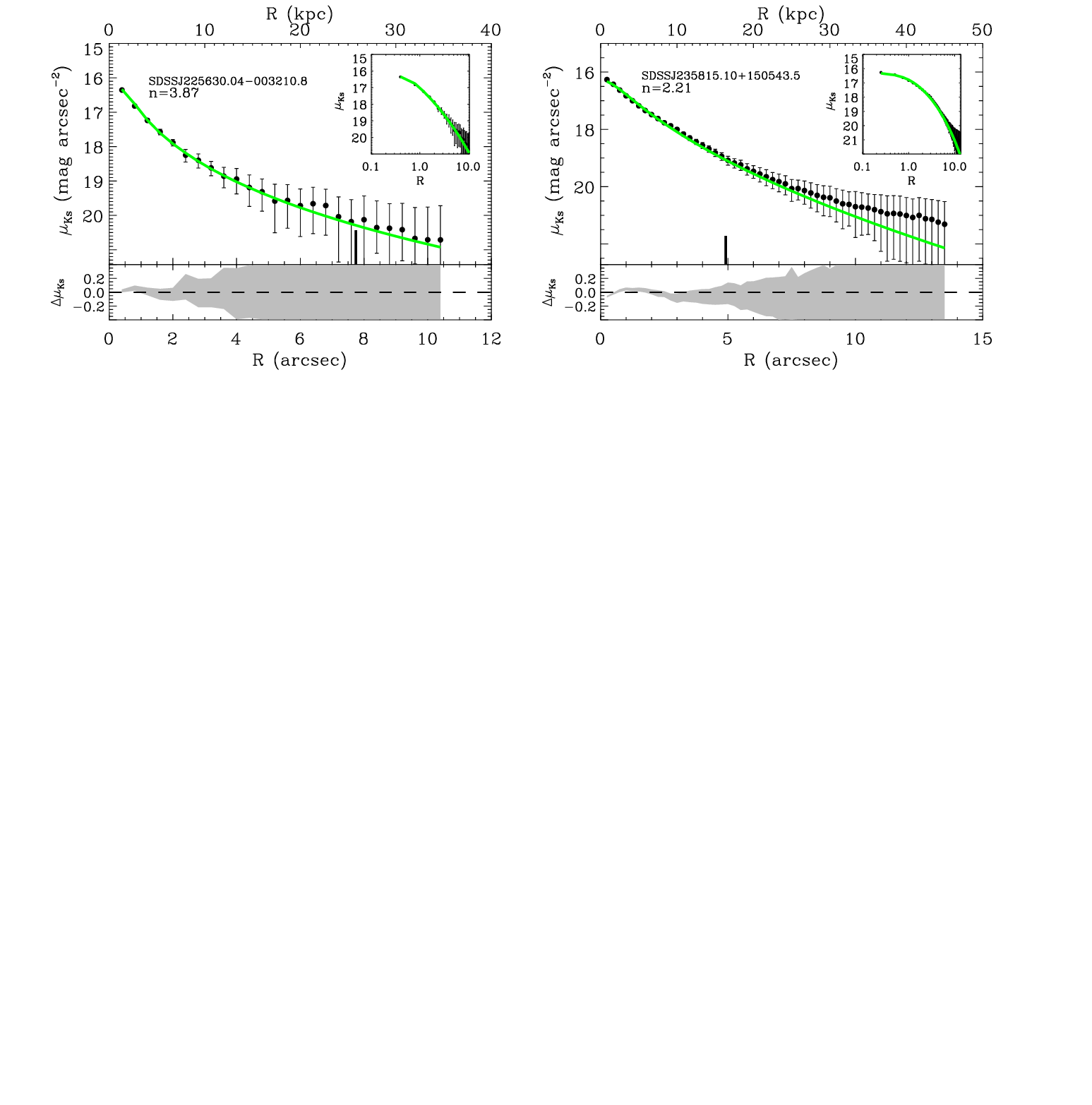}}%
  \caption[]{Continued.}
  \label{fig:SB_profiles}
\end{figure*} 
%
The  parameters derived for  the structural  components of  the sample
galaxies   using    a   S\'ersic   fit   are    collected   in   Table
\ref{tab:struc_param}.  The  values of  this table were  corrected for
galaxy   inclination,    cosmological   dimming,   K-correction,   and
evolutionary effects. K-correction  and evolutionary effect were taken
into  account  using  the  recipes  by  \citet{poggianti97}.  All  the
quantities plotted  in the figures  of this paper show  such corrected
values, unless otherwise stated.

\begin{table*}[!ht]
\caption{\label{tab:struc_param} Structural parameters in the $K_s$-band derived by fitting a S\'ersic model to our sample of BGGs}
\centering
\begin{tabular}{l c c c c c c c}
\hline\hline
\multicolumn{1}{c}{Group} &
\multicolumn{1}{c}{$\mu_{\rm e}$}&
\multicolumn{2}{c}{$r_{\rm e}$}&
\multicolumn{1}{c}{$n$}&
\multicolumn{1}{c}{$q$}&
\multicolumn{1}{c}{PA}&
\multicolumn{1}{c}{$K_s$}\\
\multicolumn{1}{c}{}&
\multicolumn{1}{c}{(mag/arcsec$^{2}$)}&
\multicolumn{1}{c}{(arcsec)}&
\multicolumn{1}{c}{(kpc)}&
\multicolumn{1}{c}{}&
\multicolumn{1}{c}{}&
\multicolumn{1}{c}{($^{\circ}$)}&
\multicolumn{1}{c}{}\\
\multicolumn{1}{c}{(1)}&
\multicolumn{1}{c}{(2)}&
\multicolumn{1}{c}{(3)}&
\multicolumn{1}{c}{(4)}&
\multicolumn{1}{c}{(5)}&
\multicolumn{1}{c}{(6)}&
\multicolumn{1}{c}{(7)}&
\multicolumn{1}{c}{(8)}\\
\hline
SDSSJ015021.27-100530.5 &  19.16$\pm$0.24 &  4.14$\pm$0.73  & 21.0$\pm$3.7  & 3.68$\pm$0.40 & 0.72$\pm$0.02 & 144.4$\pm$4.4 & 14.08$\pm$0.28 \\
SDSSJ015241.95+010025.5 &  19.50$\pm$0.12 &  6.77$\pm$0.51  & 24.9$\pm$1.9  & 2.39$\pm$0.12 & 0.52$\pm$0.01 &  18.5$\pm$0.2 & 13.12$\pm$0.10 \\
SDSSJ075244.19+455657.3 &  17.27$\pm$0.07 &  3.43$\pm$0.14  & 3.5 $\pm$0.1  & 4.56$\pm$0.10 & 0.95$\pm$0.01 & 125.4$\pm$1.4 & 11.36$\pm$0.09 \\
SDSSJ080730.75+340041.6 &  20.23$\pm$0.39 &  8.67$\pm$2.75  & 29.5$\pm$9.3  & 3.57$\pm$0.48 & 0.80$\pm$0.02 &  73.4$\pm$1.5 & 13.03$\pm$0.32 \\
SDSSJ084257.55+362159.2 &  18.39$\pm$0.42 &  3.51$\pm$1.09  & 15.0$\pm$4.6  & 3.59$\pm$0.73 & 0.82$\pm$0.02 &  89.7$\pm$1.2 & 13.41$\pm$0.33 \\
SDSSJ084449.07+425642.1 &  18.19$\pm$0.15 &  6.05$\pm$0.54  & 6.4 $\pm$0.6  & 3.84$\pm$0.21 & 0.70$\pm$0.01 &  20.9$\pm$0.1 & 11.15$\pm$0.08 \\
SDSSJ090303.18+273929.3 &  19.14$\pm$0.38 &  3.56$\pm$1.29  & 21.5$\pm$7.8  & 2.02$\pm$0.43 & 0.73$\pm$0.05 & 119.9$\pm$5.3 & 15.08$\pm$0.47 \\
SDSSJ094829.04+495506.7 &  21.00$\pm$0.42 &  8.81$\pm$3.09  & 48.0$\pm$16.8 & 5.33$\pm$0.39 & 0.45$\pm$0.01 &  53.0$\pm$1.1 & 14.22$\pm$0.34 \\
SDSSJ104302.57+005418.2 &  18.36$\pm$0.34 &  3.11$\pm$0.74  & 7.0 $\pm$1.7  & 2.88$\pm$0.48 & 0.76$\pm$0.02 & 151.0$\pm$3.0 & 13.19$\pm$0.25 \\
SDSSJ105452.03+552112.5 &  18.20$\pm$0.25 &  2.72$\pm$0.62  & 16.0$\pm$3.7  & 1.75$\pm$0.22 & 0.78$\pm$0.02 & 101.3$\pm$3.3 & 14.72$\pm$0.35 \\
SDSSJ111439.76+403735.1 &  18.89$\pm$0.21 &  2.72$\pm$0.41  & 9.0 $\pm$1.4  & 4.41$\pm$0.39 & 0.93$\pm$0.03 &  82.3$\pm$2.5 & 14.08$\pm$0.27 \\
SDSSJ112155.27+104923.2 &  18.61$\pm$0.34 &  2.92$\pm$0.79  & 11.1$\pm$3.0  & 2.71$\pm$0.41 & 0.80$\pm$0.03 & 100.9$\pm$3.0 & 14.03$\pm$0.34 \\
SDSSJ114128.29+055829.5 &  19.20$\pm$0.37 &  5.10$\pm$1.44  & 16.0$\pm$4.5  & 2.24$\pm$0.44 & 0.76$\pm$0.02 & 100.7$\pm$2.1 & 13.32$\pm$0.31 \\
SDSSJ114647.57+095228.1 &  18.95$\pm$0.15 &  4.58$\pm$0.49  & 16.3$\pm$1.7  & 2.04$\pm$0.17 & 0.54$\pm$0.01 &  55.0$\pm$1.1 & 13.47$\pm$0.19 \\
SDSSJ124742.07+413137.6 &  19.97$\pm$0.46 &  9.89$\pm$3.81  & 26.6$\pm$10.2 & 3.03$\pm$0.63 & 0.58$\pm$0.02 & 149.5$\pm$3.2 & 12.37$\pm$0.36 \\
SDSSJ133559.98-033129.1 &  18.16$\pm$0.25 &  3.53$\pm$0.55  & 10.6$\pm$1.6  & 1.99$\pm$0.30 & 0.80$\pm$0.02 & 131.1$\pm$2.6 & 13.10$\pm$0.21 \\
SDSSJ154855.85+085044.3 &  18.39$\pm$0.03 &  14.80$\pm$0.30 & 20.3$\pm$0.4  & 2.82$\pm$0.03 & 0.63$\pm$0.01 &  59.8$\pm$0.3 & 10.65$\pm$0.06 \\
SDSSJ161431.10+264350.3 &  18.39$\pm$0.08 &  6.37$\pm$0.45  & 19.7$\pm$1.4  & 2.76$\pm$0.14 & 0.78$\pm$0.02 &  20.5$\pm$0.4 & 12.93$\pm$0.18 \\
SDSSJ225630.04-003210.8 &  19.29$\pm$0.08 &  7.75$\pm$0.54  & 27.9$\pm$2.0  & 3.87$\pm$0.19 & 0.68$\pm$0.01 & 128.5$\pm$2.6 & 13.37$\pm$0.18 \\
SDSSJ235815.10+150543.5 &  18.91$\pm$0.33 &  4.91$\pm$1.18  & 14.8$\pm$3.6  & 2.21$\pm$0.39 & 0.69$\pm$0.01 &  97.2$\pm$1.9 & 13.08$\pm$0.27 \\
\hline
\end{tabular}
\tablefoot{Col.(1):   Group  name;   Col.    (2):  effective   surface
  brightness;  Col.   (3):   effective  radius  (arcsec);  Col.   (4):
  effective radius (kpc); Col.   (5): S\'ersic index; Col.  (6): minor
  to major axis ratio; Col  (7): position angle; Col.  (8): $K_s$-band
  total magnitude  derived from  the S\'ersic fit.   Effective surface
  brightness  and magnitudes  were corrected  for  galaxy inclination,
  cosmological dimming, K-correction and evolutionary effects.}
\end{table*}


\subsection{Internal errors in the structural parameters}

The errors  given in Table \ref{tab:struc_param}  for every structural
parameter were  obtained through a series of  Monte Carlo simulations.
Due  to the fact that formal  errors obtained  from  the $\chi^2$  minimization
method   are   usually  not   representative   of   the  real   errors
\citep{mendezabreu08},  we   carried  out  extensive   simulations  on
artificial galaxies  in order to  give a reliable estimation  of these
errors. Monte Carlo simulations have also the advantage of getting rid
of  systematic  biases  in  our  measurements, as  for  example  those
introduced    by   the    effects    of   seeing    in   our    images
\citep[see][]{trujillo01}.

A set of  2000 images of galaxies modeled with  a S\'ersic profile was
generated.  The structural parameters  of the artificial galaxies were
randomly chosen among the following ranges:
\begin{eqnarray}
1   &\leq& r_{\rm e} \leq 20 \,{\rm kpc}; \qquad 0.5 \leq q \leq 0.9 \nonumber\\
0.5 &\leq& n \leq 6; \qquad \qquad 10 \leq K_{s} \leq 14 \,{\rm mag}
\end{eqnarray}

In order to  mimic the same instrumental setup,  we added a background
level  and photon  noise to  these  artificial images  similar to  the
observed images.  They were  also convolved simulating the seeing that
we have  in our observations.  Finally, these  simulated galaxies were
fitted with  the same  conditions as the  real ones.   Then, simulated
galaxies were  used to determine  the errors of the  fitted structural
parameters.  To assign to  every single galaxy the corresponding error
for every structural parameter, we divided our catalogue of artificial
galaxies in bins of 0.5 mag,  we assumed that the errors were normally
distributed,  with mean  and standard  deviation corresponding  to the
systematic  and  typical  error,  respectively.  Then,  we  placed  our
observed galaxy in  its magnitude bin and assigned  to every parameter
the corresponding error.

\subsection{Radial profiles}

\begin{figure*}[!t]
   \centering
   \includegraphics[width=0.8\textwidth]{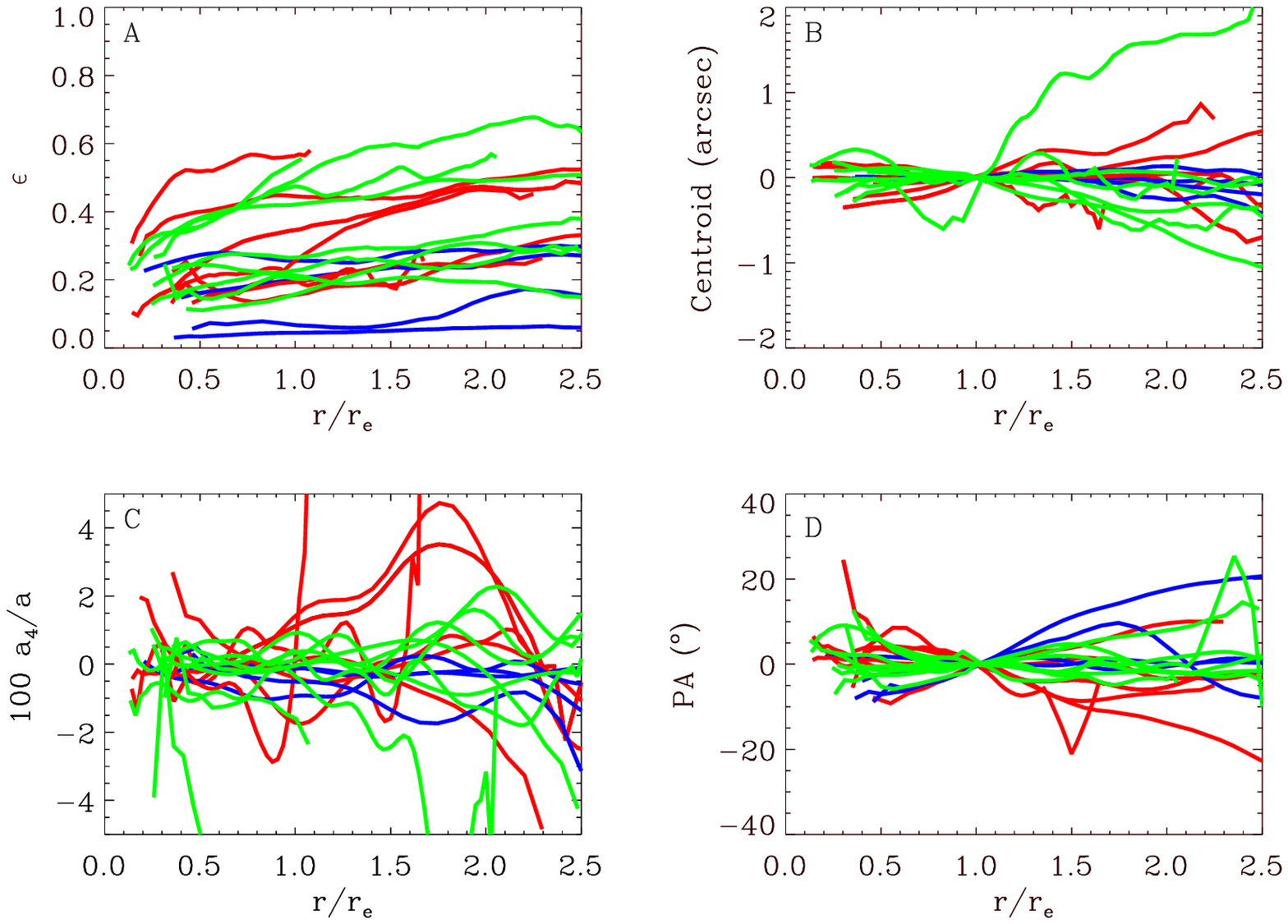}
   \caption{(A)   Ellipticity,  (B)  centroid   ($r=\sqrt{(x-x_0)^2  +
       (y-y_0)^2}$), (C)  fourth order Fourier cosine  moment, and (D)
     position  angle  radial profiles  of  our  sample galaxies.   All
     radial  profiles are normalized  to the  effective radius  in the
     abscissa axis.   Blue, green and  red solid lines refers  to BGGs
     with       $M_{K_{s}}\geq-26$,      $-26<M_{K_{s}}<-27$,      and
     $M_{K_{s}}\leq-27$, respectively.  Both the centroid and position
     angle profile were reported to  their values at $r_{\rm e}$.  The
     radial   profiles  are   plotted  for   radii  larger   than  the
     corresponding PSF FWHM.}
              \label{fig:radial_profile}%
    \end{figure*}

The  surface  photometry  was  measured  by fitting  ellipses  to  the
isophotes of  the galaxies  using the {\rm  IRAF} task  {\tt ELLIPSE}.
This routine fits a large  number of free parameters to every isophote
using the  iterative method described  by \citet{jedrzejewski87}.  Bad
pixels,  foreground  stars, and  other  spurious  sources were  masked
before performing the  fit.  Figure~\ref{fig:radial_profile} shows the
ellipticity ($\epsilon$), centroid  position ($x_{\rm 0}, y_{\rm 0}$),
fourth order  cosine Fourier moment  ($a_4$), and position  angle (PA)
radial  profiles of  the isophotes.   It  is worth  noticing that  the
surface-brightness  profiles of  16  galaxies extend  at  least to  $2
r_{\rm  e}$ at  a level  of $\mu_{K_S}  \approx 21$  mag arcsec$^{-2}$
corresponding to $1\sigma$ of the sky  counts, and 9 reach at least $3
r_{\rm e}$, denoting thus the depth of our images.

No  galaxies  with  ellipticities  larger  than  0.7  were  found,  in
agreement with  typical ellipticals (Fig.  \ref{fig:radial_profile}A).
We found  that in almost  all cases the ellipticity  profile increases
with radius  up to large distances.   We checked that  this effect was
neither caused  by a  bad sky subtraction  nor by  light contamination
from a close  object.  This behaviour cannot be  explained by assuming
that these  galaxies are oblate  (or prolate) spheroids  and therefore
suggests the presence of  another component or a structural distortion
in  the  outer  parts  \citep{porter91,mendezabreu10}.   There  is  no
correlation with the S\'ersic shape parameter, whereas we found a weak
dependence  of  the  ellipticity  with  the  galaxy  magnitude.   Less
luminous galaxies are rounder.

The variations of the galaxy centroid do not show any correlation with
the magnitude  (Fig.  \ref{fig:radial_profile}B).  Nevertheless, these
variations seem to  be larger for radius greater  than $r_{\rm e}$ at
least  for   some  galaxies,  possibly  indicating   the  presence  of
distortions in the outer parts of these galaxies.

The  fourth  order cosine  Fourier  coefficient  is  related with  the
isophotes  shape.  Negative  values indicate  boxy isophotes.   On the
contrary, positive  values are related with disky  isophotes.  For our
sample galaxies these profiles are  too noisy and this prevented us to
assign    a    single    value,     or    even    define    a    trend
(Fig.~\ref{fig:radial_profile}C).   Therefore, no  obvious correlation
was  found between  this  coefficient and  either  the S\'ersic  shape
parameter or galaxy magnitude.

The      position      angle      profiles      are      shown      in
Fig.  \ref{fig:radial_profile}D.  We  did  not  find  any  correlation
between these profiles and  other galaxy parameters.  In addition, the
profiles  are  very  flat  with  variations  $\mid  {\rm  PA}  -  {\rm
  PA}(r_{\rm e})\mid < 10^{\circ}$.

\section{Scaling relations}
\label{sec:scaling}
We studied the structural scaling  relations for our sample of BGGs in
FGs  by  analyzing  the  results from  the  photometric  decomposition
presented in Sect.  \ref{sec:photdec}.   We also considered the sample
of early-type galaxies  by \citet[][hereafter P98]{pahre98}, described
in detail in \citet{pahre99}, as  a control sample.  This represents a
statistically significant  sample of  galaxies in clusters  and groups
observed homogeneously in the $K_s$ band.  We selected the P98 BCGs to
compare the same  type of objects. The P98 BCGs  showed in the figures
are the followings: NGC~4874  and NGC~4889 (Coma), NGC~545 and NGC~547
(Abell  194), NGC~6166 (Abell  2199), NGC~7720  (Abell~2634), NGC~4696
(Centaurus),  NGC~1316   (Fornax),  NGC~3309  and   NGC~3311  (Hydra),
NGC~1272  and  NGC~1275  (Perseus),  NGC~4486  (Virgo),  and  NGC~1407
(Eridanus).

\subsection{Consistency with  P98 data}

The P98 data represents a large sample of early-type galaxies observed
in the near-infrared  with a photometric depth similar  to that of our
images. Therefore, it  is ideal to compare with our  sample of BGGs in
fossil  systems.  However,  their  structural parameters  ($L_{K_{s}}$,
$r_{\rm  e}$,  and  $\langle\mu_{\rm  e}\rangle$)  were  derived  from
circular  aperture  photometry  which  differs from  our  approach  of
performing two-dimensional photometric decompositions.

\begin{figure*}[!t]
   \centering
   \includegraphics[width=0.8\textwidth]{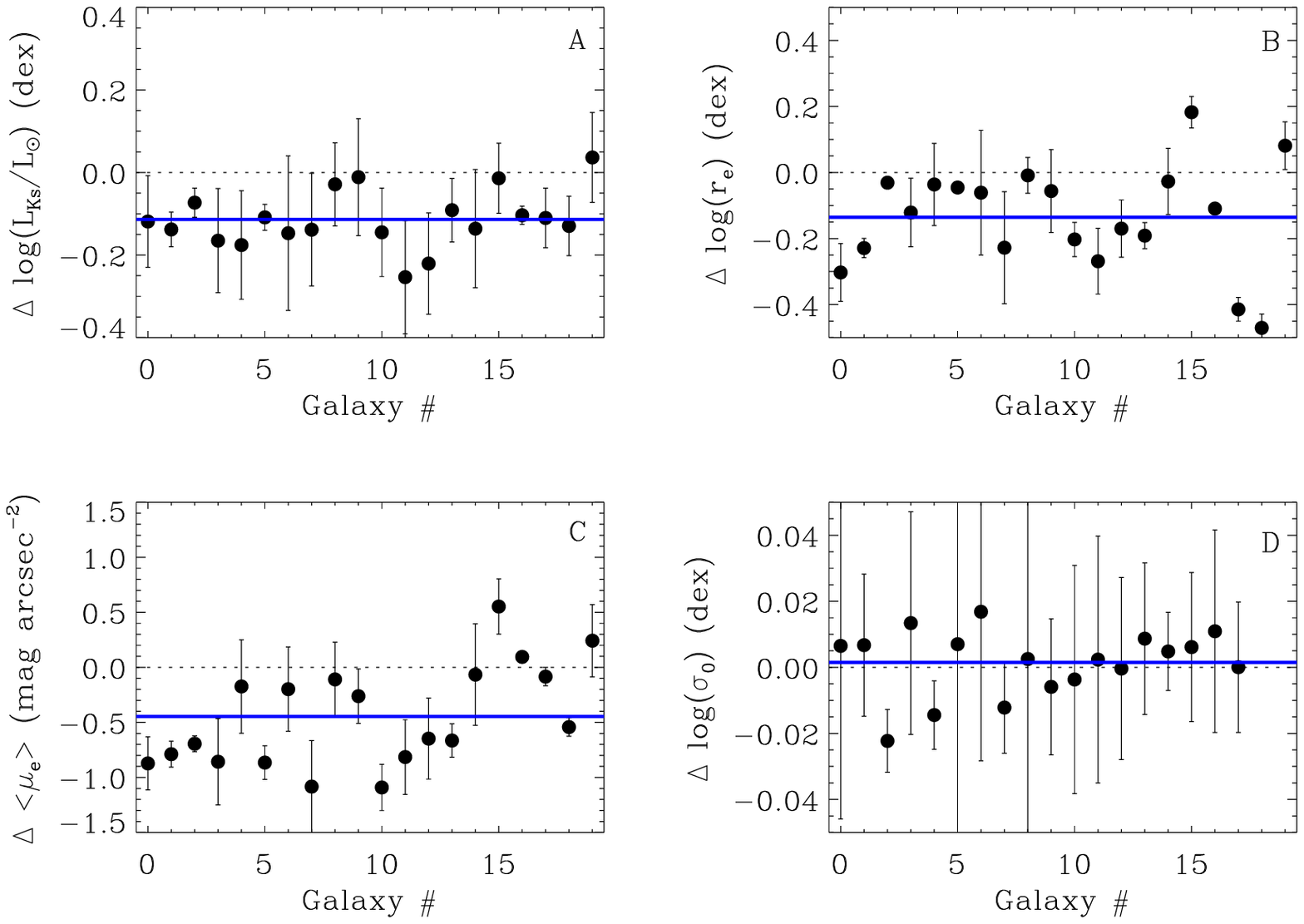}
   \caption{(A) luminosity , (B)  effective radius, (C) mean effective
     surface-brightness,  and   (D)  velocity  dispersion  differences
     between  the aperture  photometry  and photometric  decomposition
     measurements. Abscissa  axes represent a number  assigned to each
     galaxy. Blue solid line represent the mean difference taking into
     account  all points. Error  bars represent  the errors  listed in
     Tables \ref{tab:observations} and \ref{tab:struc_param}.}
              \label{fig:comparison}%
    \end{figure*}

In order  to check for possible  biases between the  two datasets that
could contaminate our results,  we calculate the structural parameters
for our sample  data using also circular aperture  photometry onto our
images. The actual  computation was done using the  {\rm IRAF} package
{\tt ELLIPSE},  as done by P98. Figure  \ref{fig:comparison} shows the
results from this  comparison.  A clear bias  appears in all the
photometric  parameters under  study.   The luminosity  ($L_{k_{s}}$),
effective radius ($r_{\rm  e}$), and mean effective surface-brightness
($\langle\mu_{\rm e}\rangle$) derived by the photometric decomposition
are 0.11 dex, 0.13 dex ,  and 0.44 mag arcsec$^{-2}$ larger than those
derived by  aperture photometry, respectively.   These differences are
not  unexpected and  have  been  pointed out  in  other studies.   For
instance,  \citet{aguerri05} and \citet{trujilloaguerri04}  found that
the differences in magnitudes between  the two methods is never larger
that 0.5 mag, in agreement with  our result. The main reason for these
differences  is  that models  used  in  the photometric  decomposition
approach  are extrapolated  to  the infinite,  therefore given  larger
magnitudes, effective radius and effective surface-brightness.

Figure  \ref{fig:comparison}  also shows  the  comparison between  the
central velocity dispersion  of the galaxies using P98  and this work.
As  explained in  Sect.~\ref{sec:spectro}, we  corrected  our velocity
dispersions  using a physical  aperture of  $r_{\rm e}$/8  whereas P98
used  a  physical  scale   of  1.53  kpc.   These  different  aperture
corrections do  not introduce  a bias between  the two samples  and it
only could increase the scatter in the relations.

These biases  must be kept in mind  in the following but  they are not
enough to change the results presented  in this work. In fact, most of
the differences are  within the error bars of  the measurements.  
  To avoid confusions,  we have corrected for these  biases in all the
  figures where we compare our measurements with P98 data.

\subsection{Structural parameters vs. luminosity}

\begin{figure}[!ht]
\centering
\includegraphics[width=0.49\textwidth,bb=105 360 530 700]{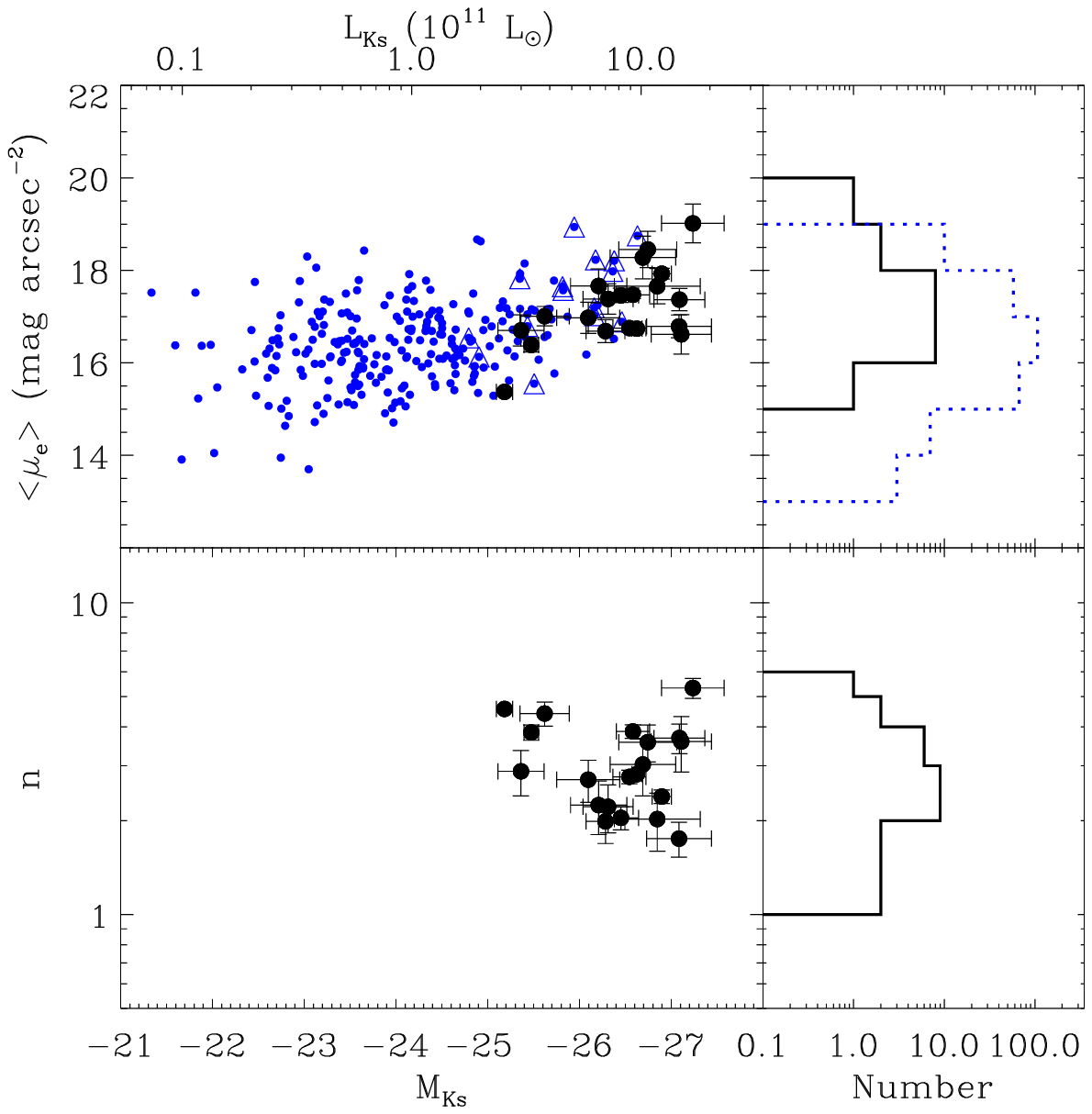}
\caption{Correlations   between  $\langle\mu_{\rm   e}\rangle$  (upper
  panel), $n$ (bottom panel) and  the galaxy absolute magnitude in the
  $K_s$-band.   Black large  dots indicate  our sample  galaxies. Blue
  dots  represent the  P98 early-type  galaxies. The  BCGs in  the P98
  sample are marked by blue open triangles.}
\label{fig:sample_mag}
\end{figure}

Figure~\ref{fig:sample_mag}  shows the  relation between  the absolute
rest-frame   magnitude  in  $K_s$-band   (or  equivalently   also  the
$K_s$-band luminosity  in solar units) obtained from  the S\'ersic fit
to the surface-brightness distribution  of the sample galaxies and the
mean effective surface brightness, or S\'ersic shape parameter.

Notice that  some of our sample galaxies  are among the brightest
  galaxies  located in  cluster/group environments  in  the Universe.
They are  brighter than  the BCGs present  in nearby clusters  such as
Virgo or  Coma, which are  represented by triangle  symbols.  However,
they are neither the most concentrated, nor posses the highest central
surface  brightness. In  fact, the  mean effective  surface brightness
correlates well with  the galaxy magnitude and follows  the same trend
of normal ellipticals and BCGs.

The P98  sample does  not provide information  about the shape  of the
surface-brightness profile,  which in our  case is represented  by the
S\'ersic index,  and therefore  we can not  make a  direct comparison.
However, it is worth noticing  the wide range of S\'ersic index values
obtained for  our BGGs, ranging from 1.7  to 5.3 with a  mean value of
3.0. This  indicates that the shape of  the surface-brightness profile
can be very different from a typical de Vaucouleurs profile.

\subsection{Fundamental plane}
\label{sec:FP}
It is  well known that  the structural parameters of  classical bulges
and elliptical galaxies follow a tight relation called the Fundamental
Plane \citep[FP; ][]{djorgovskidavis87,dressler87} among the effective
radius   ($\log  r_{\rm  e}$),   mean  effective   surface  brightness
($\langle\mu_{\rm e}\rangle$) and  central stellar velocity dispersion
($\log \sigma_0$).  The small intrinsic dispersion of the FP relation is
a consequence of the virial theorem and demonstrate that these systems
are  in dynamical equilibrium.   Possible deviations  from the  FP are
related  to  a  variation   of  the  mass-luminosity  ratio  with  the
luminosity and/or  a non-homology in  the structure of  these galaxies
throughout the whole range of luminosities \citep{trujillo04}.  The FP
has  been  widely  used  as  a powerful  tool  in  measuring  galactic
distances  and  also in  studies  of  galaxy  formation and  evolution
\citep{kjargaard93,jorgensen96,jorgensen99,kelson97}.

\begin{figure*}[!ht]
\centering
\includegraphics[width=0.8\textwidth]{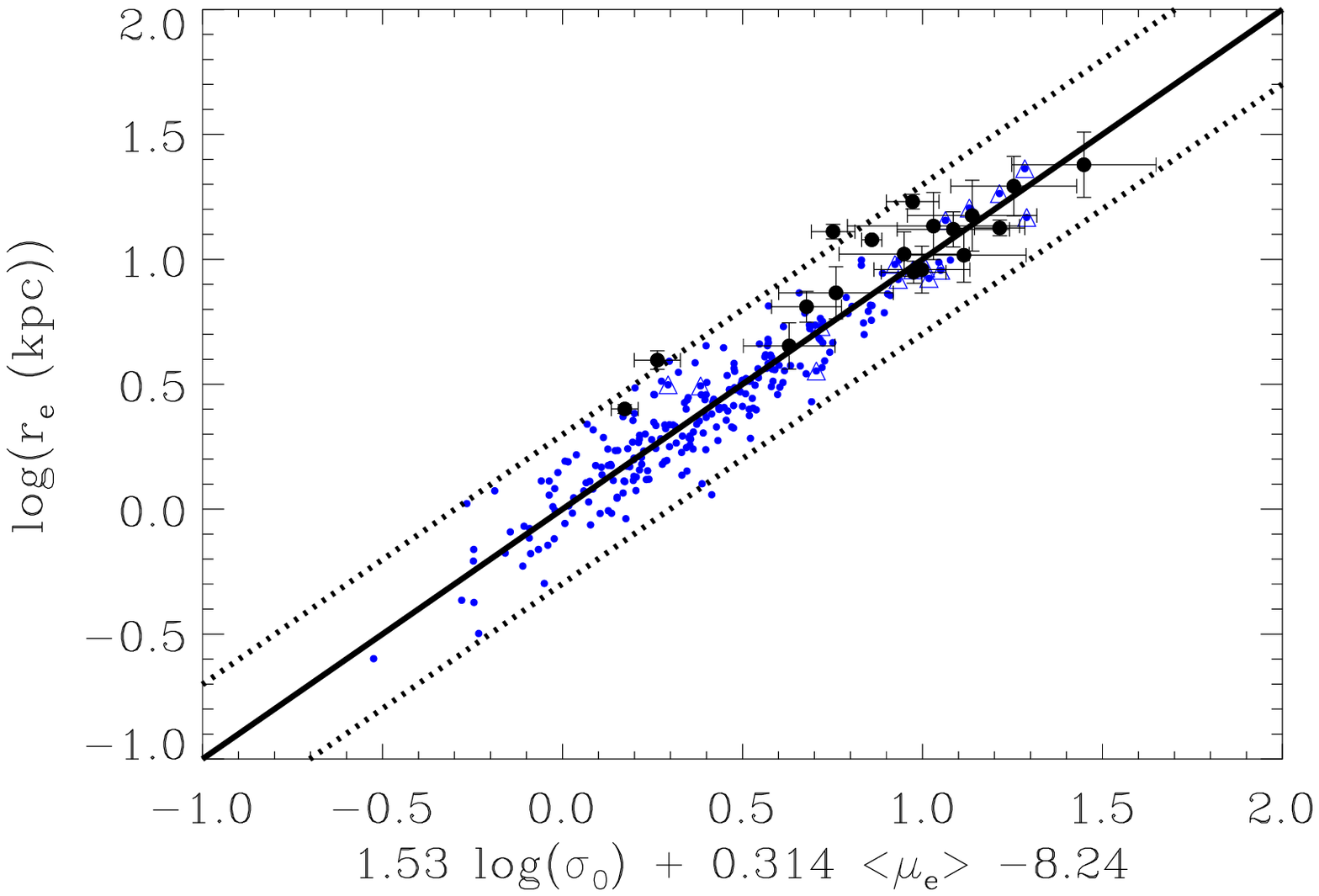}
\caption{Distribution of our BGGs  (large black points) in the edge-on
  view of  the FP defined by  the P98 early-type  galaxies (small blue
  points).  The  BCGs  of the  P98  sample  are  marked by  blue  open
  triangles.   The  solid line  represents  the  FP  best fit  to  P98
  galaxies. The  dotted lines represent the $3  \sigma$ deviation from
  the fit.}
\label{fig:FP}%
\end{figure*}

Figure~\ref{fig:FP} shows  the distribution of our  sample galaxies in
the FP  relation defined by  the sample of  P98. Our BGGs follow  a FP
compatible within the errors with that of normal ellipticals and BCGs,
therefore implying that in  the innermost regions ($r<r_{\rm e}$) they
are relaxed systems similar to normal ellipticals.

\subsection{Kormendy relation}
\label{sec:kr}

The  effective  radius  and  mean  surface  brightness  of  elliptical
galaxies  are  correlated  through  the  so-called  Kormendy  relation
\citep[KR; ][]{kormendy77} which represents a projection of the FP.

\begin{figure*}[!ht]
\centering
\includegraphics[width=0.8\textwidth]{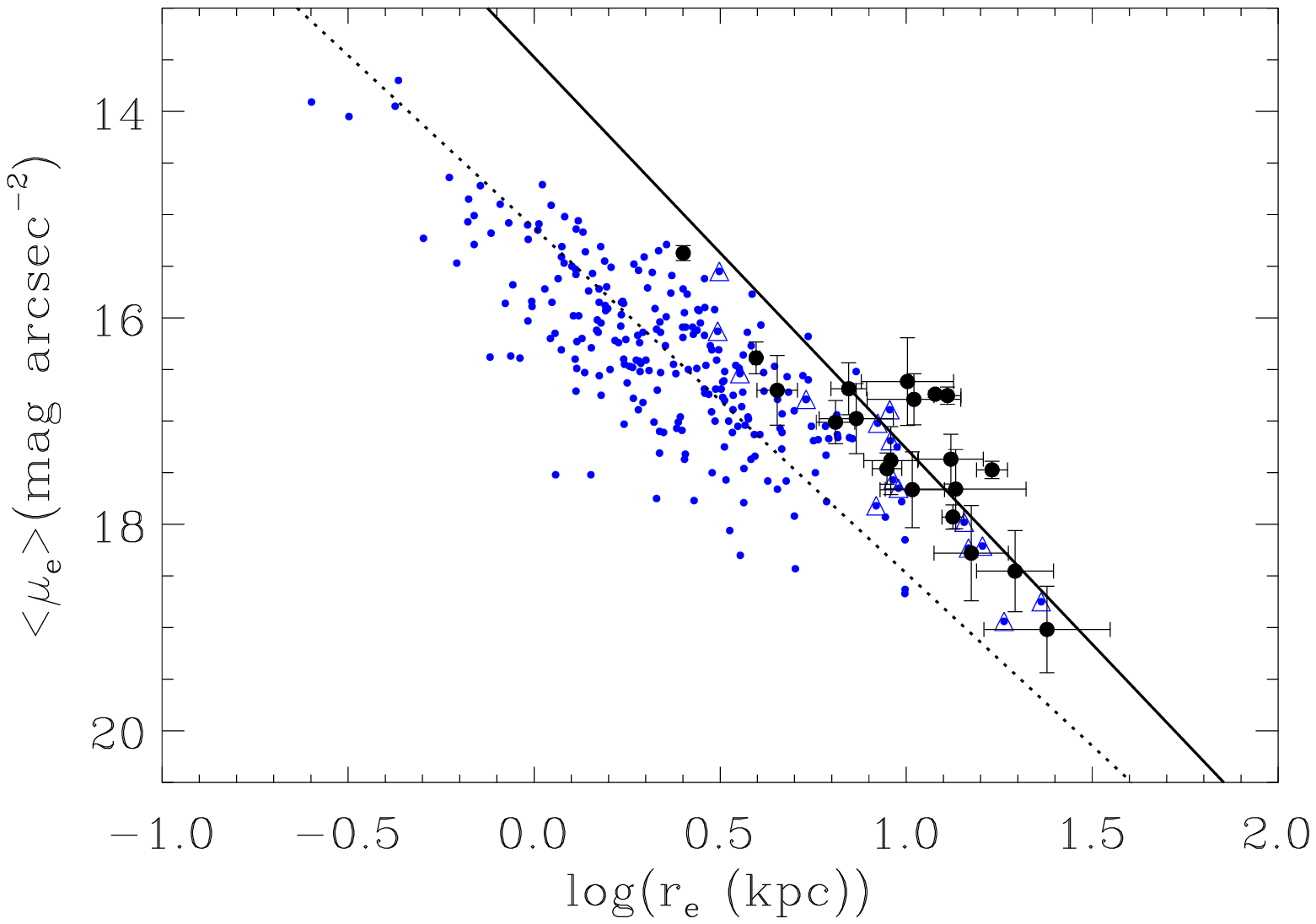}
\caption{The KR for  our BGGs (large black points)  and P98 early-type
  galaxies (small blue points).  The BCGs of the P98 sample are marked
  by blue open  triangles.  The solid line represents  the KR best fit
  to our sample and the dotted  line represents the best KR fit to P98
  sample.}
\label{fig:KR}
\end{figure*}

Figure~\ref{fig:KR}  shows the  position of  our BGGs  in the  KR with
respect to the  control sample of P98. The fit to  the data points was
done   using  the  FITEXY   routine  \citep{press92}   implemented  in
IDL\footnote{Interactive  Data Language is  distributed by  ITT Visual
  Information  Solutions.},  which   estimates  the  parameters  of  a
straight-line  fit  taking  into  account  for  data  errors  on  both
variables.  The best fitted KR to the P98 sample is given by

\begin{equation}
\langle\mu_{\rm e}\rangle = 15.13(\pm0.38) + 3.34(\pm0.86) \log r_{\rm e}.
\end{equation}

Our galaxies  are systematically above the  best-fit relation obtained
for the P98 sample, even if both relations are similar once the errors
are taken  into account.   The relation obtained  by fitting  only our
galaxies is

\begin{equation}
\langle\mu_{\rm e}\rangle =  13.48(\pm2.71) + 3.78(\pm1.51) \log r_{\rm e}.
\end{equation}

Several works studied the KR in galaxy clusters demonstrating that the
intrinsic  dispersion  is   approximately  0.4  mag  arcsec$^{-2}$  in
$\langle               \mu_{\rm               e}              \rangle$
\citep{hoessel87,sandageperelmuter91,labarbera03}.       This     high
intrinsic dispersion can be interpreted in different ways: by the fact
that  the  KR does  not  consider the  third  parameter  of FP  (i.e.,
velocity dispersion,  \citealt{ziegler99}); because of  the measurement
errors  as  well as  the  systematic  errors  due to  the  photometric
calibration and  also due to the corrections  introduced for different
biases (e.g.,  zero point  and color transformation,  K-correction and
reddening); or because the position  of the galaxies in the KR depends
on their magnitude \citep{nigochenetro08}.

In the case of our BGGs,  we suggest that since our galaxies represent
the  brightest  galaxies  located  in  cluster/groups,  the  different
position  of the  BGGs in  the  KR is  produced only  by their  higher
luminosity.  In  fact, galaxies with increasing  magnitudes are placed
parallel   to  the   KR   but   with  higher   values   of  the   mean
surface brightness for a given effective  radius, as it is the case of
our sample  galaxies.  In  addition, the coefficients  of our  fit are
poorly  constrained  due to  our  small  number  statistics, and  both
samples are consistent within the errors.

\subsection{Faber-Jackson relation}

\begin{figure*}[!ht]
\centering
\includegraphics[width=0.8\textwidth]{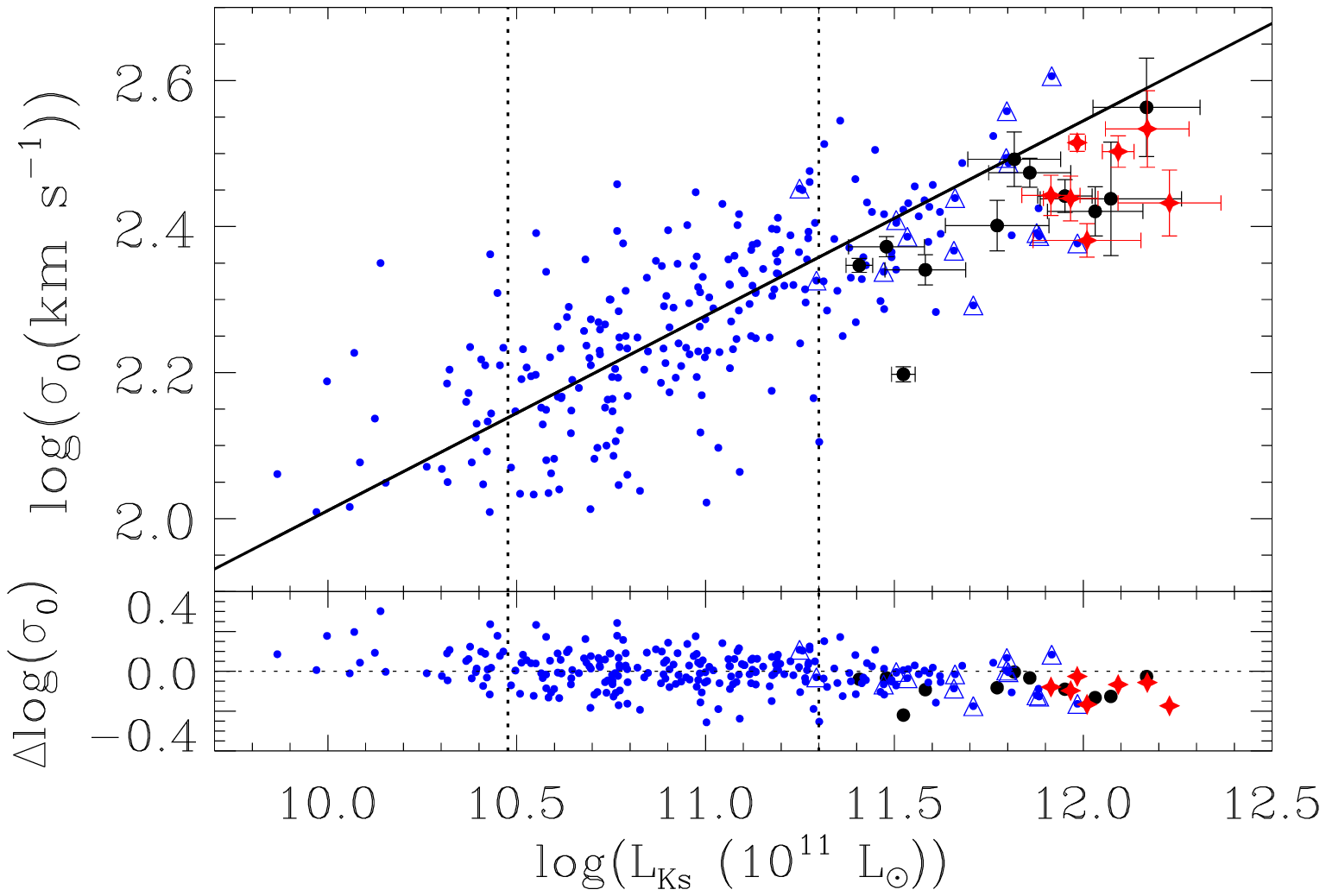}
\caption{Distribution of  our BGGs (red stars and  large black points)
  and  P98  early-type  galaxies  (small  blue points)  in  the  $\log
  \sigma_0$ vs.   $\log L_{k_{s}}$ plane.  The BCGs in the  P98 sample
  are marked  by blue  open triangles. The  solid line  represents the
  best    fit   to    the   galaxies    in   the    luminosity   range
  3$\times$10$^{10}<L_{K_{s}}/$L$_{\sun}< 2\times$10$^{11}$ as done by
  \citet{bernardi11b}.   Red stars  and large  black  points represent
  BGGs   with  ellipticities  $\epsilon>0.3$   and  $\epsilon\leq0.3$,
  respectively.  The  bottom panel  represents the residuals  from the
  best fit.}
\label{fig:FJ}%
\end{figure*}

The  Faber-Jackson relation \citep[FJ;  ][]{faberjackson76} represents
another physically  significant projection  of the FP.   It correlates
the galaxy magnitude with the stellar velocity dispersion and for this
reason  it is  observationally more  expensive to  obtain than  the KR
since it requires for spectroscopic  data. For the sake of clarity, in
Fig.~\ref{fig:FJ}  we  plot the  total  luminosity  in the  $K_s$-band
instead  of the  magnitude.  The  $K_s$-band luminosities  of  a given
stellar population are less affected by metallicity effects than those
in optical passbands. In addition, they  are a good proxy of the total
stellar  mass  since  the   typical  $M/L\sim1$  for  an  old  stellar
population \citep[e.g.,][]{bruzualcharlot03}.

Figure~\ref{fig:FJ} represents the FJ  relation for the P98 sample and
our  sample of  BGGs.  Recently,  \citet{bernardi11a,bernardi11b} have
demonstrated that the scaling relations of early-type galaxies present
a change in their slope  at a characteristic stellar mass ($M_* \sim2
\times10^{11}$ M$_{\sun}$,      see     also     \citealt{tortora09}).
Interestingly, this curvature of  the scaling relations is not present
when  the stellar  mass of  the galaxies  is replaced  by  the central
velocity  dispersion.  In  particular,  they find  that galaxies  with
masses  higher  than   2  $\times$10$^{11}$  M$_{\sun}$  have  smaller
velocity dispersions and larger effective radius for a given mass than
their   counterparts   in    the   mass   range   3$\times$10$^{10}$$<
M_*/$M$_{\sun}<$ 2$\times$10$^{11}$.

Our sample galaxies represent a subset of the most massive galaxies of
the Universe.   Some of  them probably fall  in the  Bernardi's sample
\citep{bernardi11a,bernardi11b}      since     they      have     SDSS
information. Unfortunately, we cannot compare directly with their work
since  they  have optical  photometry  whereas  we have  near-infrared
photometry.  In Fig.  \ref{fig:FJ} we plot  the FJ fit to the range of
masses  where  \citet{bernardi11b}   consider  that  the  relation  is
linear. Most of  our BGGs follows the expected  curvature in the $\log
\sigma_0$ vs.  $\log L_{K_{s}}$ relation, confirming and extending the
results by \citet{bernardi11b}.

\subsection{Effective radius vs. luminosity}

\begin{figure*}[!ht]
\centering
\includegraphics[width=0.8\textwidth]{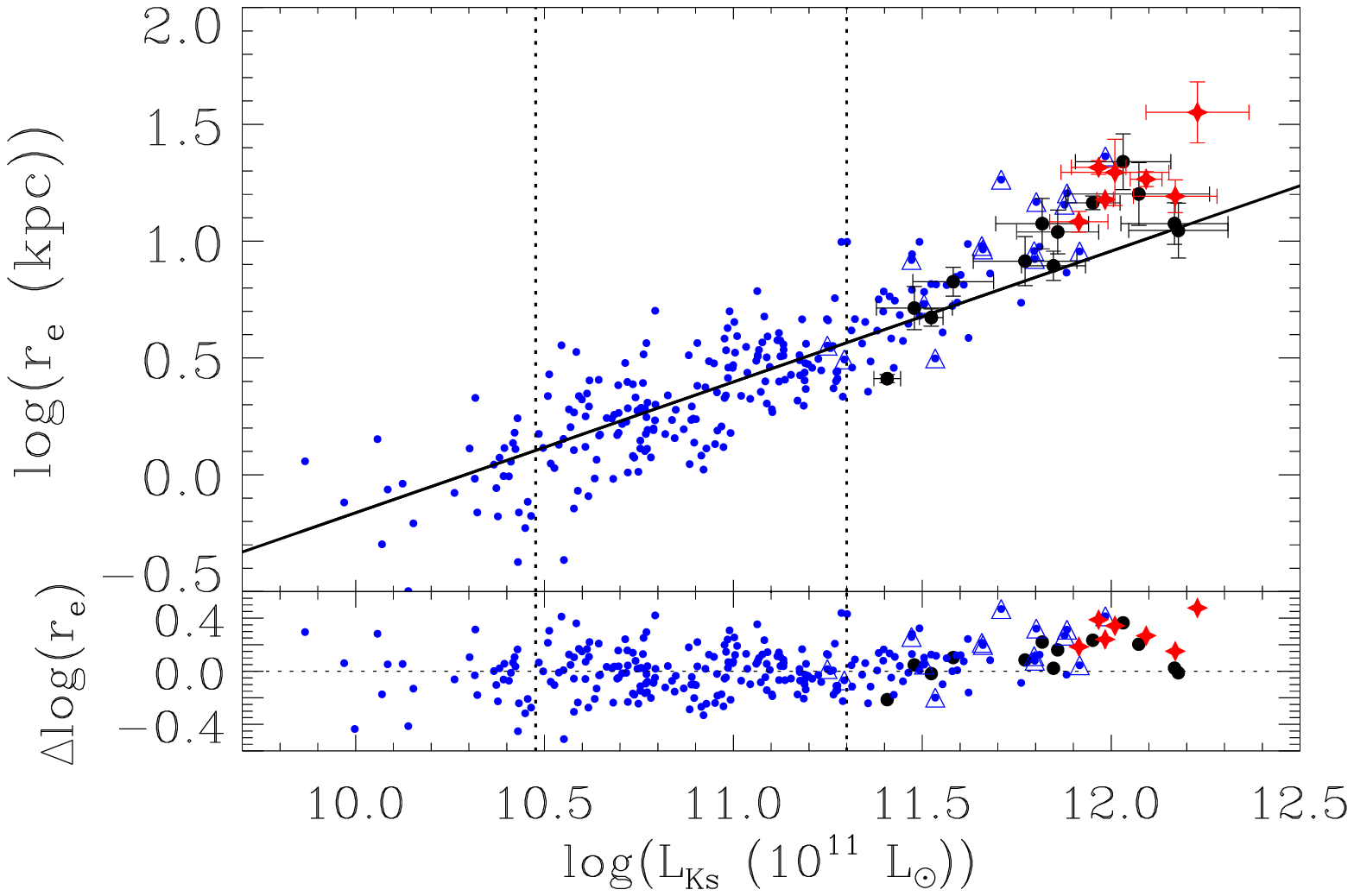}
\caption{Distribution of  our BGGs (red stars and  large black points)
  and P98 early-type galaxies (small  blue points) in the $\log r_{\rm
    e}$ vs.  $\log L_{K_{s}}$ plane.  The  BCGs in the  P98 sample are
  marked by  blue open triangles.  The solid line represents  the best
  fit     to    the     galaxies    in     the     luminosity    range
  3$\times$10$^{10}<L_{K_{s}}/$L$_{\sun}< 2\times$10$^{11}$ as done by
  \citet{bernardi11b}.   Red stars  and large  black  points represent
  BGGs   with  ellipticities  $\epsilon>0.3$   and  $\epsilon\leq0.3$,
  respectively.  The  bottom panel  represents the residuals  from the
  best fit.}
\label{fig:bernardi}%
\end{figure*}

Figure~\ref{fig:bernardi}  shows the  relation between  the $K_s$-band
luminosity  in solar  units  obtained  from the  S\'ersic  fit to  the
surface brightness and the effective radius of the galaxies.

As  done   in  the  previous   section,  we  tested  the   results  of
\citet{bernardi11b}  with our  sample of  massive BGGs.   We  fitted a
straight  line in  logarithmic scale  to the  data in  the  mass range
3$\times$10$^{10}<M_*/$M$_{\sun}<$2$\times$10$^{11}$  and extrapolated
such a fit to larger masses.  From  our data, it is clear that most of
our BGGs  follow the  expected curvature in  the $\log r_{\rm  e}$ vs.
$\log L_{K_{s}}$  relation, i.e., galaxies  with masses higher  than 2
$\times$10$^{11} $M$_{\sun}$ have larger  effective radius for a given
mass   than   their   counterparts   in   the   mass   range   between
3$\times$10$^{10}<M_*/M_{\sun}<2\times$10$^{11}$,  in  agreement  with
\citet{bernardi11b}.

\section{Discussion}
\label{sec:discussion}

The BGGs  of the  fossil systems presented  in this study  represent a
subset of  the brightest  and most massive  galaxies in  the Universe.
This can be seen by  looking at both their $K_s$-band luminosities and
central stellar velocity dispersions. However, they closely follow the
FP scaling  relation defined by  normal elliptical galaxies  and other
BCGs, suggesting  that they are dynamically relaxed  systems at least
in their central regions.

FGs  are believed to  be created  fast and  in an  early epoch  of the
Universe by  the efficient  merger of $L^*$  galaxies inside  DM halos
\citep{donghia05,diazgimenez11}.  In this  scenario, the BGGs observed
at low  redshift had time to  reach a virialized  status although some
remnant of their  merger phase could still be  present.  The different
relaxation time scales between the  inner and outer regions imply that
the outskirts of galaxies may have fossil information about their past
formation.  The  continuously increasing ellipticity  profile observed
in almost all galaxies of our sample  could be a hint of such a merger
history.    \citet{porter91}  reported  on   this  behaviour   of  the
ellipticity profiles in a sample of BCGs, they found that ~50\% of the
BCGs  in rich  clusters present  these characteristics  and associated
them  with the observed  absence of  rotation in  luminous ellipticals
\citep{davies83,carter85}.  This  result might be  consistent with our
findings if  the increasing  ellipticity profiles are  correlated with
galaxy luminosity.   In addition, \citet{boylankolchin06} demonstrated
by using  numerical simulations that  these ellipticity trends  can be
explained in terms of a high fraction of radial collisions, indicating
a preferential direction in the orbits of the mergers.

Assuming    that    FGs    were    formed   in    an    early    epoch
\citep{dariush07,diazgimenez11} and then  passively evolved for a long
time in  order to reach a  relaxed state, it is  customary to consider
that  dissipational mergers  with a  high  fraction of  gas should  be
predominant in  the formation of  these systems \citep{diazgimenez08}.
Recent  works   by  \citet{robertson06}  and   \citet{hopkins08}  have
demonstrated through  simulations and observations,  respectively, how
the observed tilt of the FP can be explained in terms of the amount of
dissipation involved  in the  formation of elliptical  galaxies.  They
conclude   that  galaxies   formed  through   dissipationless  mergers
perfectly  follow the  FP defined  by the  virial theorem  ($r_{\rm e}
\propto   \sigma_{0}^{2}  \langle   \mu_{\rm   e}  \rangle^{0.4}$)   whereas
ellipticals  formed through  dissipational mergers  will present  a FP
tilt depending  on the  fraction of gas  involved in the  merger.  Our
galaxies, as  normal ellipticals, are  well described by a  tilted FP,
thus showing some indications that  most of them could be the remnants
of  mergers  with  a   high  gas  fraction.   In  addition,  numerical
simulations  show that  gas-rich mergers  at  an early  stage of  the
formation of BGGs are needed to account for the observed properties of
elliptical galaxies \citep{hernquist93,naabostriker09}.

However, numerical  simulations of gas-rich mergers  also predict that
the  remnants will  be round  and compact  galaxies (i.e.,  with small
effective  radius)  where  the  stellar  velocity  dispersions  should
increase   \citep[e.g.,][]{khochfarsilk06,naab07,hopkins08}.    Recent
observational     works     on      early     type     galaxies     by
\citet{bernardi11a,bernardi11b} have  shown a  change in the  slope of
different scaling relations that involve the stellar mass.  This trend
is interpreted  by the authors  in terms of  the kind of  mergers that
have  been  involved  in   the  formation  of  these  systems.   Major
dissipationless  mergers between galaxies  are mechanisms  expected to
increase the  final size of the  resulting galaxy but  not its central
velocity dispersion; however, if minor dry mergers are the predominant
mechanism,  they are  expected to  change both  the size  and velocity
dispersion    \citep{bernardi11a}.    Therefore,   \citet{bernardi11b}
claimed that the most feasible scenario is that galaxies with $M_* > 2
\times$10$^{11}  $M$_{\sun}$ formed  mainly  by major  dissipationless
mergers.   It is  clear  that most  of  our BGGs  follow the  expected
curvature  in   both  the   $\log  \sigma_0$  vs.    $\log  L_{K_{s}}$
(Fig.~\ref{fig:FJ})  and  $\log   r_{\rm  e}$  vs.   $\log  L_{K_{s}}$
(Fig.~\ref{fig:bernardi})  relations,   confirming  the  observational
trend found  by \citet{bernardi11b}.  This  interpretation contradicts
the previous one hinting toward a formation scenario where most of our
BGGs formed through major gas-rich merger \citep{khosroshahi06}.

Another commonly used proxy  to understand the formation mechanisms of
early-type  galaxies  is the  S\'ersic  index.  Numerical  simulations
predict S\'ersic indexes for the  remnants of dry mergers in the range
$3<n<8$  \citep{gonzalezgarciabalcells05,naabtrujillo06},  whereas  in
dissipational   mergers   they  would   be   in   the  range   $2<n<4$
\citep{hopkins09,kormendy09}.  We have  tentatively divided our sample
in galaxies with $n\leq3$ and $n>$3 in order to understand whether the
shape of the surface-brightness  profiles could be used to distinguish
among the different formation scenarios.  We did not find any relation
between the departure from the intermediate-mass scaling relations and
S\'ersic   index.   Similarly,  several   works  based   on  numerical
simulations found that the shape of  the isophotes can be also used to
trace the merger  history of elliptical galaxies \citep{naabburkert03,
  khochfarburkert05,naab06}.   These  studies  found  out  that  disky
isophotes are  the result of  gas-rich mergers whereas  boxy isophotes
are preferentially formed in dissipationless collisions.  Supported by
these results, \citet{khosroshahi06} claim that BGGs in fossil systems
were  formed through  dissipational mergers  since they  show non-boxy
isophotes.   We measured  the isophote  shapes by  means of  the $a_4$
coefficient,      however,     their     noisy      profiles     (Fig.
\ref{fig:radial_profile})  prevented us  to extract  conclusions about
the merger history of our BGGs based on the isophotes shape.

Interestingly, we  found that the ellipticity turned  out to be a
  good indicator  of the kind of  merger involved in  the formation of
  BGGs. Numerical simulations show  that the remnants of dissipational
  mergers  are  rounder  on  average  when  compared  to  remnants  of
  dissipationless  mergers \citep{cox04}.   In  fact, the  ellipticity
  distribution  of dissipational  merger remnants  peaks at  about 0.2
  which  is in sharp  contrast to  the dissipationless  remnants whose
  distribution  of ellipticities  peaks at  about 0.4.   We  used this
  information to separate our  BGGs in galaxies with $\epsilon\leq0.3$
  and $\epsilon >  0.3$. We measured the ellipticities  at $r_{\rm e}$
  to mimic  simulations. The results are shown  in Figs.  \ref{fig:FJ}
  and  \ref{fig:bernardi}.    It  is  clear  that   BGGs  with  larger
  ellipticities  systematically  deviate  from  the  intermediate-mass
  scaling relation, whereas rounder galaxies are also distributed near
  to this relation.  Therefore, we  interpret this result as a further
  hint that  dissipationless mergers ($\epsilon  > 0.3$) are  the main
  processes driven the evolution of BGGs.

{\it  We   conclude  that  BGGs  in  fossil   systems  suffered  major
  dissipational mergers in an early  epoch of their formation, but the
  bulk of their mass is assembled latter in subsequent dissipationless
  mergers that increase the BGGs size}.

\section{Conclusions}
\label{sec:conclusions}

We analyzed  the near-infrared $K_s$-band photometric  properties of a
sample of  20 BGGs observed within  the framework of  the FOGO project
\citep{aguerri11}. So far, this represent the  largest sample of
  BGGs  in  fossil systems  studied  in  the  near-infrared. Here  we
summarize our results:

\begin{itemize}

\item The  structural parameters of  the sample galaxies  were derived
  using  the  GASP2D  algorithm  \citep{mendezabreu08}  by  fitting  a
  S\'ersic  model to the  their surface-brightness  distribution. This
  model provides a good description  for all the galaxies and allows a
  straightforward comparison with the results available in literature.

\item Only one  sample BGG is a  candidate to be a cD  galaxy since it
  shows an upward break in the surface-brightness profile with respect
  to a de Vaucouleurs profile.

\item The ellipticity profile of most of our galaxies is an increasing
  function of  the radius.  This can  be related to the  fact that our
  galaxies   represent  the   very  bright   end  of   the  luminosity
  function. The  trend in the ellipticity profile  cannot be explained
  in terms  of a oblate/prolate spheroid. Together  with the variation
  observed in the galaxy  centroid and position angle radial profiles,
  it  might indicate  the presence  of preferential  direction  in the
  orbits of the mergers.

\item We built  the $K_s$-band scaling relations (FP,  KR, and FJ) for
  our  sample  of BGGs  and  we compared  them  with  those of  normal
  elliptical  and BCGs.  We  did not  find any  significant difference
  between  the  FP  defined  by  BGGs  and  those  of  ellipticals  or
  BCGs. This  suggests that  the central regions  of BGGs  are relaxed
  with structure similar to  those of normal ellipticals.  However, we
  found a change  of the slope in  the FJ and in the  $\log r_{\rm e}$
  vs.  $\log L_{K_{s}}$ relation for our massive galaxies.
\end{itemize}

These  observational  results  can  be  interpreted in  terms  of  the
formation scenario  of the  BGGs.  Our BGGs  sample follows the  FP of
ellipticals  and  therefore  are  well  reproduced  by  a  tilted  FP.
Therefore,  this supports  that  most of  our  BGGs probably  suffered
dissipational mergers during  their formation.  However, the curvature
found in the  scaling relation involving the mass  of the galaxies can
be interpreted as an indication that the stellar mass of these systems
grew mainly  by  dissipationless mergers. Therefore,  we suggest
that BGGs in fossil system  suffered major dissipational mergers in an
early  epoch  of  their formation,  but  the  bulk  of their  mass  is
assembled latter  in subsequent dissipationless  mergers that increase
the BGGs size.

\begin{acknowledgements}

We acknowledge  the anonymous referee for the  positive comments which
helped  us to  improve the  paper.  JMA,  JIP, and  JVM  are partially
funded by the Spanish MICINN under the Consolider-Ingenio 2010 Program
grant     CSD2006-00070:     First     Science    with     the     GTC
(http://www.iac.es/consolider-ingenio-gtc).    JMA    and   JALA   are
partially  funded by the  Spanish MICINN  (grants AYA2007-67965-C03-01
and AYA2010-21887-C04-04).   JIP and JVM  are partially funded  by the
Spanish MICINN (grants AYA2007-67965-C03-02 and AYA2010-21887-C04-01).
EMC is supported by the University of Padua through grants CPDA089220,
60A02-1283/10, and 60A02-5052/11 and by the Italian Space Agency (ASI)
through  grant   ASI-INAF  I/009/10/0.   This  article   is  based  on
observations made with the  William Herschel Telescope operated on the
island  of La  Palma, in  the Spanish  Observatorio del  Roque  de los
Muchachos of the Instituto de Astrof\'isica de Canarias.  This work is
based in part on data obtained  as part of the UKIRT Infrared Deep Sky
Survey.

\end{acknowledgements}


\bibliographystyle{aa} 
\bibliography{reference} 


\end{document}